\documentclass[aps,prl,amsmath,amsfonts,amssymb,notitlepage,twocolumn,superscriptaddress,footinbib,floatfix]{revtex4-2}
\newtheorem{theorem}{Theorem}

\usepackage[T1]{fontenc}
\usepackage[utf8]{inputenc}

\usepackage{CJK}
\usepackage{physics} 
\usepackage{soul}
\usepackage[caption=false,subrefformat=parens,labelformat=parens]{subfig}
\usepackage{color} 
\usepackage[table,dvipsnames]{xcolor}
\usepackage[colorlinks,linktocpage,bookmarks=false]{hyperref} 
\usepackage[normalem]{ulem}   

\usepackage{tikz}
\usepackage{tikz-cd} %for \sout

\newcommand\myshade{85}
\definecolor{myrulecolor}{RGB}{150,20,0}% define the color for the rules
\colorlet{mylinkcolor}{violet}
\colorlet{mycitecolor}{YellowOrange}
\colorlet{myurlcolor}{Aquamarine}
\hypersetup{
	linkcolor  = myurlcolor!\myshade!black,
	citecolor  = mycitecolor!\myshade!black,
	urlcolor   =  myrulecolor!\myshade!black,
}

\usepackage{graphicx}
\graphicspath{{FIG/}}
\usepackage{multirow} 
\usepackage{booktabs} % book-quality tables

\AtBeginDocument{
	\heavyrulewidth=.08em
	\lightrulewidth=.05em
	\cmidrulewidth=.03em
	\belowrulesep=.65ex
	\belowbottomsep=0pt
	\aboverulesep=.4ex
	\abovetopsep=0pt
	\cmidrulesep=\doublerulesep
	\cmidrulekern=.5em
	\defaultaddspace=.5em
}% needed to make booktabs work with revtex

\newcommand{\Z}{\mathbb{Z}}

\newcommand{\prlsection}[1]{\noindent\textbf{{#1}--- }}
\renewcommand\[{\begin{equation}}
\renewcommand\]{\end{equation}}
\makeindex 
\begin{document}  
\begin{CJK*}{UTF8}{gbsn} % Use default fonts from CJK (see below)
    \title{Generalized Kramers-Wannier Duality from Bilinear Phase Map}
    \author{Linhao Li}  
   \affiliation{Department of Physics, the Pennsylvania State University, University Park, Pennsylvania 16802, USA}
    \affiliation{Department of Physics and Astronomy, Ghent University, Krijgslaan 281, S9, B-9000 Ghent, Belgium}

    \author{Masaki Oshikawa}  
    \affiliation{Institute for Solid State Physics, The University of Tokyo. Kashiwa, Chiba 277-8581, Japan}

	\author{Han Yan (闫寒)}   
    %\thanks{The two authors contributed equally to this work}
    \email{hanyan@issp.u-tokyo.ac.jp}
    \thanks{Corresponding author.}
    \affiliation{Institute for Solid State Physics, The University of Tokyo. Kashiwa, Chiba 277-8581, Japan}

	\date{\today}
    \begin{abstract}
We present the bilinear phase map (BPM), a concept that generalizes the Kramers-Wannier (KW) transformation to investigate unconventional gapped phases of qudit spin chains. Encoding the transformation in a matrix, the BPM  enables the exploration of a broader spectrum of generalized quantum phases and dualities, including the nonunitarity in duality transformations, and derivation of general noninvertible fusion rules. 
Furthermore, we  
obtain strong constraints on anomaly conditions of qudit models with a general class of duality symmetries. This establishes a microscopic anomaly classification for a broad class of lattice dualities.
    \end{abstract}
\maketitle 

\end{CJK*}

\prlsection{Introduction   }
Identifying distinct quantum phases of matter and understanding phase transitions between them stands as a central challenge in quantum many-body physics. Recent decades have witnessed the discovery of a multitude of exotic gapped phases, e.g., symmetry protected topological (SPT) phase \cite{PhysRevB.80.155131, PhysRevB.81.064439, PhysRevB.85.075125, PhysRevB.87.155114}, topological orders \cite{Wen1995,kitaev2003fault,PhysRevLett.96.110405,PhysRevLett.96.110404}, fracton orders \cite{PhysRevLett.94.040402,PhysRevB.92.235136,PhysRevB.94.235157} and spontaneous symmetry breaking (SSB) phases. Intriguingly, some gapped phases are interconnected to each other through duality transformations, even though they exhibit vastly different physical properties. The most well-known example is Kramers-Wannier (KW) transformation \cite{Kramers-Wannier}, which relates the paramagnetic phase and ferromagnetic phases of the transverse field Ising chain \cite{PhysRev.60.252,Shankar_2017,cobanera2011bond,RevModPhys.51.659,cobanera2011bond}. As these two phases have different ground state degeneracies, the KW transformation is realized by a nonunitary operator, which satisfies the non-invertible ``Ising-category'' fusion rule \cite{PhysRevLett.93.070601,Frohlich:2006ch,Bhardwaj:2017xup,Aasen:2016dop,Aasen:2020jwb,PhysRevLett.126.195701,PhysRevB.104.125418,10.21468/SciPostPhys.11.4.082,PRXQuantum.4.020357,PhysRevB.108.214429,Moradi:2022lqp,Moradi:2023dan,Cao:2023doz}. The loss of unitarity can be recovered by introducing symmetry twisted boundary conditions \cite{PhysRevB.108.214429,Cao:2023doz,Lootens:2022avn,PhysRevB.98.155137,PhysRevLett.126.217201}. Moreover, when the system is self-dual, the KW duality is promoted to an anomalous noninvertible symmetry \cite{PhysRevResearch.2.043086,Kong:2020jne,Frohlich:2009gb,PhysRevLett.128.111601,Kaidi:2022cpf,Kaidi:2022uux,PhysRevD.105.125016,Choi:2022jqy,PhysRevLett.130.131602,Choi:2022zal,Cordova:2022ieu,Bhardwaj:2022yxj,Bhardwaj:2022lsg,Bartsch:2022mpm,Bhardwaj:2022kot,Bartsch:2022ytj,Bhardwaj:2023fca,Inamura:2023ldn,Bhardwaj:2022maz,Huang:2023pyk,Chatterjee:2022jll,Chang:2022hud,Delcamp:2023kew,Putrov:2023jqi,Sun:2023xxv,Pace:2023mdo,Fechisin:2023dkj,Sinha:2023hum,Shao:2023gho,Choi:2023vgk,Pace:2023kyi,Inamura:2023qzl,Chen:2023czk,Fukusumi:2023vjm,AksoySciPostPhys2024}. This anomaly forbids gapped phases with a unique
ground state (which we shall call uniquely gapped phases for short), including the SPT phases
\cite{Chang:2018iay,Thorngren:2019iar,Thorngren:2021yso,Zhang:2023wlu,Cordova:2023bja,Apte:2022xtu,Seiberg:2024gek,Antinucci:2023ezl,Nagoya:2023zky,Cao:2024qjj}. Thus the self-dual point must be a first-order or continuous phase transition between the duality-related phases.  
Despite these advances, a structural classification of dualities on the lattice --particularly a microscopic criterion for when a duality is anomalous -- remains lacking.

To address this problem, we formulate them as bilinear phase maps (BPMs), whose linear-algebraic properties encode invertibility, symmetry, and anomaly data.
In this perspective, noninvertibility and various key properties of the KW duality naturally arise from the rank deficiency of the linear map,
or equivalently the existence of nontrivial kernels. 
For more general BPMs, the nontrivial kernels, which dictate the corresponding dualities, are determined by the characteristic polynomials.  

Furthermore, we construct a family of quantum $p$-state spin (prime-$p$-level qudit) model  %with neighboring 3-site interactions,
, which is subject to a generalized three-site BPM duality.
Most importantly, we prove the following theorem for such a broad class of self-dual quantum models: the combined symmetry of three-site BPM duality and reflection is anomalous (i.e., forbids a uniquely gapped phase) if and only if $-1$ is not quadratic residue modulo $p$. 
And as a corollary, the smallest anomalous case occurs at $p=3$, revealing new physics absent in qubit systems.

\prlsection{The Kramers-Wannier Duality  }
We first review the  KW duality of the spin-1/2 (qubit) chains,  as a preparation for discussion of the generalized KW duality.
    
We consider a closed chain with $L$ sites. On each site $i$ sits a spin-1/2 variable $s_i\in\{0,1\}$. We also consider the $\Z_2$ symmetry generated by $U=\prod_j X_j$, which flips all spins, namely $s_j\to s_j+1$.
The KW transformation is realized by gauging the $\Z_2$ symmetry for the entire Hilbert space. On the one dimensional lattice, the $\Z_2$ gauge field is defined as dual spins $\{\hat{s}_{i-\frac12}\}$ on the link. The spins $\{s_i\}$ on the original lattice are mapped to dual spins $\{\hat{s}_{i-\frac12}\}$ under KW transformation. In addition,  we use $(-1)^{\hat{u}}$ to denote the eigenvalue of the dual symmetry $\hat{U}:= \prod_{i=1}^{L}\hat{X}_{i-\frac{1}{2}}$, and $\hat{t}$ to denote the boundary condition ${\hat{s}_{i-\frac12+L}}={\hat{s}_{i-\frac12}+\hat{t}}$. 

The KW transformation is realized by a mapping operator $\mathcal N$ defined as,
\begin{align}\label{eq:KWact1d}
    \begin{split}
        &\mathcal N\ket{\{s_{i}\}} \\
       & =\frac{1}{2^{\frac{L}{2}}} \sum_{\{\hat{s}_{i+\frac12}\}}(-1)^{\sum_{j=1}^{L}(s_{j-1}+s_{j})\hat{s}_{j-\frac12}+\hat{t}s_L}\ket{\{\hat{s}_{i+\frac12}\}}. 
           \end{split}    
\end{align}
The exponents in Eq.~\eqref{eq:KWact1d} are reminiscent of the minimal coupling of the gauge fields. The boundary terms in the exponents are chosen to give the correct mapping of symmetry-twist sectors.  

The KW duality  is particularly useful in understanding the physics of 1D spin chains with the same global $\mathbb{Z}_2$ symmetry, such as the quantum transverse-field lsing Hamiltonian 
$\mathcal{H}  =- \sum_{i} Z_i Z_{i+1} - h \sum_i X_i .$ 
Such KW duality exchanges these two terms, namely $h\to 1/h$, and maps the paramagnetic phase at $h>1$ to the ferromagnetic phase at $0<h<1$, which  determines the phase transition point at $h=1$ \cite{Kramers-Wannier,Shankar_2017,cobanera2011bond,RevModPhys.51.659}.\\

\prlsection{KW Duality from the Bilinear Phase Map  }
We now introduce the concept of the bilinear phase map, which will be the core of the generalized KW duality.

Under periodic boundary condition,  namely $\hat{t}=0$, Eq.~\eqref{eq:KWact1d} can be written in a more compact form after shifting $\hat{s}_{j-\frac12}$ to $\hat{s}_{j}$
\[
\label{eqn.bpm}
\begin{split}
         \mathcal N\ket{\{s_{i}\}} &=\frac{1}{2^{\frac{L}{2}}} \sum_{\{\hat{s}_{i}\}}(-1)^{\sum_{j=1}^{L}(s_{j-1}+s_{j})\hat{s}_{j-1}}\ket{\{\hat{s}_{i}\}}\,\\
         &\equiv \frac{1}{2^{\frac{L}{2}}} \sum_{\{\hat{s}_{i}\}}(-1)^{\sum_{j,k=1}^{L}s_j A_{jk}\hat{s}_{k}}\ket{\{\hat{s}_{i}\}}
\end{split}
\]
where the matrix $\vb*{A}$ is a $\mathbb{Z}_2$ valued  $L\times L$ matrix, 
\begin{equation} 
    \vb*{A}=\left(\begin{array}{ccccc}1 & 1 &0&\cdots& 0\\ 0 & 1&1&0&\cdots\\ 0&0&1&1&\cdots\\ \cdots&\cdots&\cdots&\cdots&\cdots\\1&0&\cdots&0&1
    \end{array}\right). 
\label{eq:A-for-KW_Ising}
\end{equation}
Its crucial feature  is that the rank of $\vb*{A}^{\text{T}}$ is $L-1$, and has a nontrivial kernel 
\begin{equation}
    \vb*{a} \equiv  \ker \vb*{A}  = \ker \vb*{A}^{\text{T}} = \left( 1, 1, \cdots, 1 \right). 
\end{equation}
The kernel is the root of several key properties of the KW duality. 
First, we have 
\[
\begin{split}
   &\mathcal N\ket{\{s_{i}\}}=\frac{1}{2^{\frac{L}{2}}} \sum_{\{\hat{s}_{i}\}}(-1)^{\sum_{j,k=1}^{L}s_j A_{jk}\hat{s}_{k}}\ket{\{\hat{s}_{i}\}}\\
   &= \frac{1}{2^{\frac{L}{2}}} \sum_{\{\hat{s}_{i}\}}(-1)^{\sum_{j,k=1}^{L}(s_j+a_j) A_{jk}\hat{s}_{k}}\ket{\{\hat{s}_{i} \}} \\
   &=\mathcal N\ket{\{s_{i}+1\}}.  
\end{split}
\]
That is, the duality mapping does not distinguish 
$\ket{\{s_{i}\}}$ and $\ket{\{s_{i}+ 1 \}}$: both states are mapped to the same state of the dual spins $\hat{s}_{i}$. 
Therefore, for a system with an SSB phase with two ground states $\ket{\{s_{i}\}}$ and $\ket{\{s_{i}+ 1 \}}$, the KW duality will map them to the same state. 
For the same reason, the KW duality maps the states in the odd sector  of the  $\Z_2$ symmetry (i.e., $\ket{\{s_i\}} - \ket{\{s_i + 1\}}$)
to zero. Hence, \textit{the nontrivial kernel is a sufficient condition} to the loss of unitarity of the KW duality.

Similarly, we also have 
\[
\begin{split}
&\frac{1}{2^{\frac{L}{2}}} \sum_{\{\hat{s}_{i}\}}(-1)^{\sum_{j,k=1}^{L}s_j A_{jk}\hat{s}_{k}}\ket{\{\hat{s}_{i}\}}\\ & = \frac{1}{2^{\frac{L}{2}}} \sum_{\{\hat{s}_{i}\}}(-1)^{\sum_{j,k=1}^{L}s_j A_{jk}\hat{s}_{k}}\ket{\{\hat{s}_{i}+ 1 \}}.
\end{split}
\]
That is, \textit{the state mapped to by the KW duality is always in the even sector of dual $\Z_2$ operation defined by the kernel  $\hat{s}_i\to \hat{s}_i+a_i$}.

Another feature is that the kernel 
determines the 
states  that are  mapped into the paramagnetic state    $\ket{ \rightarrow \rightarrow \dots}$. 
By definition of the mapping,  the state  $\ket{\vb*{0}} \equiv \ket{\{s_i = 0\}}$ is always mapped into $\ket{ \rightarrow \rightarrow \dots}$ 
because  $\sum_{j,k=1}^{L}s_j A_{jk} = 0$. 
It   then follows that any other states
$\ket{\{0 + a_i\}} = \ket{\{1, 1, \dots \}}$    are also mapped to the paramagnetic state.  

Finally, the twisted boundary condition terms $s_L\hat{t}$ can be understood in the context of the BPM too. 
It is simply making the replacement 
\[
\sum_{j,k=1}^{L}s_j A_{jk}\hat{s}_{k} \longrightarrow \sum_{j,k=1}^{L}s_j A_{jk}\hat{s}_{k} + \vb*{s}\cdot \hat{\vb*{t}},
\label{eq:twisted-bc_BPM}
\]
where $ \hat{\vb*{t}} = (0,\dots, 0, 1)$ for the twisted boundary condition. 
The reason why this term works is that it distinguishes the $\mathbb{Z}_2$ dual two states   $\ket{\{s_i\}} $ and  $\ket{\{s_i + a_i\}}$, i.e., $ \hat{\vb*{t}} \cdot {\vb*{s}} \ne  \hat{\vb*{t}} \cdot( {\vb*{s}}+ {\vb*{a}} )$, so the problem of $\vb*{A}$ being rank $L-1$ and hence the map being nonunitary is resolved. 
Based on this, we can actually introduce other general  $ \hat{\vb*{t}}$ with odd number of element $1$ that achieves the same purpose. Physically, such $ \hat{\vb*{t}}$'s  correspond to twisting the spins odd times on the chain. 
That is, \textit{the kernel defines the twisted boundary condition that recovers the unitarity of KW duality.}

\prlsection{Generalized Bilinear Phase Map  }
We now turn to 1+1D, $p$-level qudit systems with other types of global symmetries (here  ``global'' is defined as the symmetry operation grows linearly with the system size) and their associated KW dualities.
One such example is the symmetry of flipping only the even or odd spins  on the spin chain. We mainly consider prime number $p$ but the framework can be extended to nonprime $p$'s with more careful number theory treatment.
The spin variable $s=0,1,\ldots,p-1$ is an element of $\mathbb{Z}_p$, which is closed with respect to naturally defined
addition and multiplication (ring). When $p$ is prime, any nonzero element of $\mathbb{Z}_p$ has an inverse;
$\mathbb{Z}_p$ can thus be regarded as a field $\mathbb{F}_p$.
The BPM is now formulated in terms of linear algebra over $\mathbb{F}_p$.

For the systems whose dualities are expressed as BPMs, the properties such as the loss of unitarity and sectors of different boundary conditions \cite{PhysRevB.108.214429,Cao:2023doz,Lootens:2022avn,PhysRevB.98.155137,PhysRevLett.126.217201,kubica2018ungauging,PhysRevB.106.045125,PhysRevB.106.075150,PhysRevB.106.075150,PhysRevB.106.224420,Yao:2023bnj} can be quickly read out from the BPM:
for each generalized KW duality,
denoted as $\mathcal N_{\text{BPM}}$, and expressed the same as Eq.~\eqref{eqn.bpm} but with a different matrix $\vb*{A}$ of
$\mathbb{Z}_p$ numbers, one simply needs to examine  the linear algebraic property of
$\vb*{A}$ to straightforwardly derive these properties of the generalized duality, as summarized in the Table~\ref{Table_KW_BPM}.
\begin{table*}[ht!]
\caption{
\label{Table_KW_BPM}
Property of generalized KW duality from bilinear phase map}
\begin{tabular}{c@{\hskip 0.5in} c }
\toprule
Generalized KW duality                       & Bilinear Phase Map $\vb*{A}$ \\ 
\midrule
Nonunitary                      & Rank deficient        \\
Underlying global symmetry &  Kernel of $\vb*{A}^T$ \\
Two states map to the same state & Two states' difference is the kernel of $\vb*{A}^T$ \\
States map to paramagnet state   & Zero state  adding kernel of  $\vb*{A}^T$   \\
Boundary terms recovering unitarity  &   Linear terms differentiating kernels of  $\vb*{A}^T$ \\ 
\bottomrule
\end{tabular}
\end{table*}

Suppose the matrix  
$\vb*{A}^{\text{T}}$ have $N$ linearly independent   kernel vectors,
\begin{equation}
    \vb*{b}^m \in \ker \vb*{A}^{\text{T}} ~, m=1,\cdots,N.
\end{equation}
Then the BPM duality mapping does not distinguish state $\ket{ \{ s_{i} \} }$ and $\ket{ \{s_{i}+ n_mb^m_i \}}$, where
$s_i\in \mathbb{F}_p$ and $n_m\in \mathbb{Z}$, since
\[ \mathcal N_{\text{BPM}}\ket{\{s_{i}\}}= \mathcal N_{\text{BPM}}\ket{\{s_{i}+n_mb^m_i\}}, \forall m=1,\cdots, N,
\]
which shows explicitly that the mapping is not unitary. 
Moreover, the group $G$, which is generated by operators $U_m: \{s_{i}\} \to\{s_{i}+ b^m_i \}$ with $m=1,\cdots, N$, will be mapped to the identity group acting on the dual qudits under BPM. A similar duality mapping between Pauli operators is shown in Refs. \cite{kubica2018ungauging,tantivasadakarn2020jordan,tantivasadakarn2021long}, but from perspectives of  gauging and measurement. 
These kernels correspond to associated invertible symmetries, including the unconventional cases of dipole symmetries \cite{PhysRevB.109.125121,PhysRevB.109.115142} and exponential symmetries \cite{hu2024quantum,pace2026lieb}.  
 
This motivates us to consider the Hamiltonian invariant under $G$ since the Hamiltonian for $\hat{s}_i$ after BPM naturally commutes with the identity. 
In particular, if the $\{s_i\}$ system is in $G$-SSB phase with $p^N$ degenerate ground states (these states are $\ket{\vb*{0}}$, $\ket{\vb*{0}  + n_1\vb*{b}^1 + n_2\vb*{b}^2  + \dots }$), the dual system is in the trivial SPT phase with a unique ground state $\sum_{\{\hat{s}_{i}\}}\ket{\{\hat{s}_{i}\}}$. Moreover, an analogous analysis can be carried out for the dual $\hat{s}_i$ system and the corresponding symmetry kernels are given by
\begin{equation}
    \hat{\vb*{b}}^m \in \ker \vb*{A} ~, m=1,\cdots,N,
    \end{equation}
which also generate a $G$ group by $\hat{U}_m: \{\hat{s}_{i}\} \to\{\hat{s}_{i}+ \hat{b}^m_i \}$.
The kernel dimension directly controls whether the duality can act as a  noninvertible symmetry, and will be the key quantity entering our  anomaly classification later.

Recovering the unitarity of BPM can be achieved by additional terms $\sum^N_{m=1}\vb*{\hat{t}}_m\cdot\vb*{s}^m$ to the BPM and each term serves to distinguish the kernel state $\ket{\{s_i = {b}^m_i\}}$ and the state $\ket{\{s_i  =  0\}}$. The vector $\vb*{\hat{t}}_m$ arises from twisting the boundary conditions of the dual variables $\hat{s}_i$ using the symmetry transformation $\hat{U}_m$. Thus the number of distinct $\vb*{\hat{t}}$'s is the same as that of symmetry kernels.  
Details of how these terms are determined in concrete examples are provided in Supplemental Material~\footnote{See Supplemental Material at [URL will be inserted by publisher] for a detailed derivation of additional boundary terms of BPM in concrete examples.}.

Finally, the fusion rule of BPM and its conjugation can be directly computed by
\[
\begin{split}
&\mathcal N^{\dagger}_{\text{BPM}} \mathcal N_{\text{BPM}}\ket{\{s_{i}\}}=\sum_{g\in G}g\ket{s_i}.
\end{split}
\]
In particular, as in the case of the original KW duality of the Ising model~\eqref{eq:A-for-KW_Ising},
when the matrices $\vb*{A}$ and $\vb*{A}^T$ are related by translation over $n$ sites
\[
\begin{split}
A_{j,k}=A_{k,j+n},
 \end{split}
\]
we can further calculate the fusion of two BPMs: 
\[
\begin{split}
&\mathcal N_{\text{BPM}}  \mathcal N_{\text{BPM}}\ket{\{s_{i}\}}= T^n (\sum_{g\in G}g)\ket{s_i},
 \end{split}
\]
where $T$ is the translation operator: $T\ket{\{s_i\}}= \ket{\{s'_i=s_{i-1}\}}$ \cite{Seiberg:2023cdc,Cao:2023doz}.

Let us now explain algebraic features of the kernels  for the generalized $p$-qudit 1+1D Ising model, whose dualities are  described exactly by the generalized BPM. The  Hamiltonian is  
\[\label{eq: qudit Ising}
\mathcal{H} = \sum_j[ Z_{i}^{n_0}Z_{i+1}^{n_1} 
\dots Z_{i+k}^{n_k}  + h  X_i  + \text{H.c.}].
\]
Here the clock and shift operators acting on qudits are
\begin{equation}
Z_i=\sum_{s=0}^{p-1}\omega^s\ket{s}\bra{s}_i, \ X_i=\sum_{s=0}^{p-1}\ket{s+1}\bra{s}_i, \ \omega=e^{\frac{2\pi i}{p}}.
\end{equation}

The corresponding BPM is $A_{ji}=n_{j-i}$ when $i\le j\le i+k$ and satisfies the conditions $A_{ji}=0$ if $i>j$ or if $i+k<j$ \footnote{More precisely, under PBC, the conditions for the BPM matrix must be modified on the boundary region where $i+k> L$. In this case, the condition becomes $A_{ji}=0$ if $i+k-L<j<i$ and $A_{ji}=n_{j-i}$ if $i\le j$ and $A_{ji}=n_{j+L-i}$ if $j+L\le i+k$.}.
Thus the kernels $\vb*{b}$ of $\vb*{A}^{\text{T}}$ satisfy
\[
b_i n_0 + \dots + b_{i+k} n_k = 0
\quad \forall i.
\]
The solution of $b_i$ can be obtained by using the generating function technique. The detail of the proof is in Supplemental Material~\footnote{See Supplemental Material at [URL will be inserted by publisher] for a detailed proof of the general kernels of BPM matrix.}, and here we lay out the core results for the readers. 
We first define the characteristic polynomial to be 
\[
Q(x) = n_0 + n_1 x + \dots + n_k x^k
\]
and the general solution of $b_i$ depends on the roots of $Q(x)$.
Given a root $R$ of degeneracy $l$, there is a solution 
\[
b_i = (c_1 + c_2 i + c_3 i^2 + \dots  c_l i^{l-1}) R^i.
\]
In particular, for $p$ being a prime number, there are always $k$ solutions to the order-$k$ polynomial $Q(x) = 0$, in the field $
\mathbb{F}_p^k$ ($
\mathbb{F}_p^k$ is analogous to extending the real number field to the complex number to obtain solutions beyond real numbers). For those roots $R$'s in $
\mathbb{F}_p^k$ but not 
$\mathbb{F}_p$, proper choice of $c_i$'s makes sure that $b_i$ is still in $\mathbb{F}_p$.

The roots of $Q(x) = 0$ determine the properties of the kernel, and provide a unified view point of different spatial-modulated symmetries. 
For example, a root $R=1$ of order 1, with solution $b_i=c_1$, corresponds to a global charge symmetry, and $R=1$ of order 2, with solution $b_i=c_2 i$,  corresponds to dipole symmetry. A root $R\neq 1$ of order 1, with solution $b_i=c_1 R^i$, corresponds to exponential symmetry, and a root $R\neq 1$ of order 2,  with solution $b_i=c_2 iR^i$, corresponds to polynomial-exponential symmetry.

To concretely illustrate the physics, we have prepared a number of examples in  Supplemental Material~\footnote{See Supplemental Material at [URL will be inserted by publisher] for the detail of three-site and four-site interacting models, which includes Refs.~\cite{1992CMaPh.147..431K,PhysRevB.45.304,oshikawa1992hidden,PhysRevB.46.3486,PhysRevB.107.125158,PhysRevB.78.094404,PhysRevB.88.085114,PhysRevB.83.104411,Devakul:2018fhz,li2023intrinsicallypurely,Bhardwaj:2023bbf,ParayilMana:2024txy,chen2014symmetry,Wang:2021nrp,Li:2022jbf,PhysRevLett.133.116601}} on three-site and four-site interacting models and explored their properties in detail, including the loss of unitarity, fusion rule, duality triangle and ,etc., some of which have been discussed in previous literature too \cite{turban1982self,PhysRevB.29.2404,PhysRevB.29.519,PhysRevB.26.6334,kolb1986conformal,alcaraz1987conformal,igloi1986series,PhysRevE.90.032101,PhysRevB.108.214430}.  
More importantly, this linear-algebraic structure allows us to derive a sharp anomaly criterion for a broad class of reflection dualities, as discussed in the next section.

\prlsection{RBPM duality and its anomaly}We now reveal a strong constraint on the anomaly condition of a general class of qudit chains. 
We focus on the Hamiltonians with invertible symmetry kernels of a given three-site BPM matrix and consider the combined transformation of this three-site BPM duality and a reflection symmetry operation $R$ that maps the site $j$ to $-j$. Such Hamiltonians are compatible with this combined duality and can exhibit self-duality, since the symmetry kernel remains identical before and after this combined duality. The inclusion of reflection is needed because the dual model after BPM has symmetry kernel $\hat{b}^m_{j} = b^m_{-j}$ if $A_{ji}=n_{j-i}$. 
We dub this combined duality as the  ``reflection" BPM (RBPM):
\[
\begin{split}
   &\mathcal N_{\text{R}}\ket{\{s_{i}\}}=\frac{1}{p^{\frac{L}{2}}} \sum_{\{\hat{s}_{i}\}}\omega^{\sum_{j,k=1}^{L}s_{j} A_{jk}\hat{s}_{-k}}\ket{\{\hat{s}_{i}\}}.  
\end{split}
\] 
The three-site RBPM is defined, for $r,s \in \mathbb{F}_p$ and $r,s \neq 0$, by 
\[\label{eq:three-site RBPM}
A_{j j}=rs, A_{ j j-1}=-r-s, A_{ j j-2}=1 
\]
and the other matrix elements are zero.
The corresponding characteristic equation $(x-r)(x-s)=0$ has the solution $x=r, s \in \mathbb{F}_p$ when $p$ is prime. When $r\ne s$, two kernels are given by $b^1_j=r^j$ and $b^2_j=s^j$ while when $r=s$, two kernels are given by $b^1_j=r^j$ and $b^2_j=j r^j$.

Regardless of the values of $r$ and $s$, self-dual systems under the corresponding RBPM have the following constraint.
\setcounter{theorem}{0}
\begin{theorem}
    Such a three-site RBPM duality symmetry is anomalous if and only if
    $-1$ is not quadratic  residue modulo $p$, i.e.,
    when there exists no $k\in \Z_p$ satisfying $k^2=-1 ~\text{mod}~ p$, independent of the specific form of the duality mapping determined by $r$ and $s$. 
    In such a case, self-dual systems must have either multiple ground states or a gapless low-energy spectrum.
    In contrast, an SPT phase with a unique gapped ground state that is self-dual can exist,
    when there exists $k \in \Z_p$ such that $k^2 = -1 \mod{p}$. 
\end{theorem}  
The proof follows directly from the spectral structure of the associated BPM matrix over $\mathbb{F}_p$ and is presented in Supplemental Material~\footnote{See Supplemental Material at [URL will be inserted by publisher] for a detailed proof of the anomaly conditions of RBPM duality.}.
An immediate corollary is that the smallest anomalous prime is $p=3$, revealing new physics impossible in qubit chains. 

We now analyze the minimal anomalous realization of $p=3$. We choose $r=s=-1$. (The other examples corresponding to the remaining choices of $r$ and $s$ with $p=3$ are presented in Supplemental Material~\footnote{See Supplemental Material at [URL will be inserted by publisher] for a detailed discussion of the other minimal examples with anomalous RBPM duality symmetries, which includes Refs. \cite{lahtinen2021quantum}.}.)
Its invertible symmetry is generated by $\eta_1=\prod_j X^{(-1)^j}_j$ and $\eta_2=\prod_j X^{(-1)^j j}_j$, which we refer to as the staggered-dipole symmetry. The corresponding generalized qudit Ising model defined in Eq.\eqref{eq: qudit Ising} takes the form:
\begin{equation}\label{eq: Hal sdipole-Ising}
\begin{split}
    \mathcal{H}_{\text{sdipole-Ising}}
    &=\sum_{i=1}^{L}(\cos\theta Z_{i-1}Z_{i}^2Z_{i+1}+\sin\theta X_{i})+\text{(H.c.)},
    \end{split}
\end{equation} 
which we denote as the staggered-dipole Ising model.
The RBPM duality exchanges two terms in Eq.~\eqref{eq: Hal sdipole-Ising}, namely
$\theta \leftrightarrow \frac{\pi}{2}-\theta$.
Since the model is self-dual with respect to the anomalous RBPM duality
at $\theta=\frac{\pi}{4}$ and $\theta=\frac{5\pi}{4}$,
the system must be gapless or have degenerate ground states at these points.
Note that, in a conventional duality argument, one can argue the existence of a phase transition
at the self-dual point only with an assumption such that there is a single phase transition
between two distinct phases in the opposite sides of the self-dual point.
The anomaly of the duality has a stronger implication that the ground state
at the self-dual point cannot be trivial without any further assumptions.

Our observations based on the anomalous RBPM duality are corroborated by
existing results, as follows.
We can consider another BPM duality, which is different from the above RBPM duality,
in terms of the BPM matrix $A_{ii}=(-1)^i, A_{ii+1}=(-1)^{i+1}$.
It induces the following transformation: 
 \[
 \mathcal{N}_1: X_i\to (Z_i Z^{\dagger}_{i+1})^{(-1)^i}, \quad Z_{i-1} Z_{i}\to X^{(-1)^{i-1}}_i .
 \] 
 The staggered-dipole Ising model~\eqref{eq: Hal sdipole-Ising} is then mapped to
 \begin{equation}
    \mathcal{H}_{\text{QT}}=\sum_{i=1}^{L}(\cos\theta X_{i-1}^{\dagger}X_{i}+\sin\theta Z_{i-1}^{\dagger} Z_i)+\text{(h.c.)}.
    \label{eq:H_QT}
\end{equation}
And the RBPM duality in the staggered-dipole Ising model is mapped to the combination of reflection and the unitary transformation $U^H=\prod_j U^H_j$, where $U^H_j$ is the Hadamard gate acting on site $j$,
\[
U^{H}_j=\frac{1}{\sqrt{3}}\sum_{s,s'=0}^{2}\omega^{- ss'}\ket{s'}\bra{s}_j:\quad X_{j}\to Z_j^{\dagger},\quad Z_j\to X_{j}.
\]
This combination exchanges the two terms
in \eqref{eq:H_QT}.

Uncovering the duality between the two apparently very different models is another indication of the power of
our BPM formulation.
The latter model~\eqref{eq:H_QT} is known as the quantum torus chain, and its phase diagram
has been determined numerically in \cite{PhysRevB.86.134430,PhysRevB.104.045151}.
As the duality  preserves both the structure of the phase diagram and the central charge
of the gapless region \cite{Lootens:2022avn}, we can directly identify the phase diagram of the model \eqref{eq: Hal sdipole-Ising} as shown in Fig.~\ref{fig:phase diagram-2}, with further details provided in Supplementary Material.
\begin{figure}[htb]
 \centering
\includegraphics[width=\linewidth]{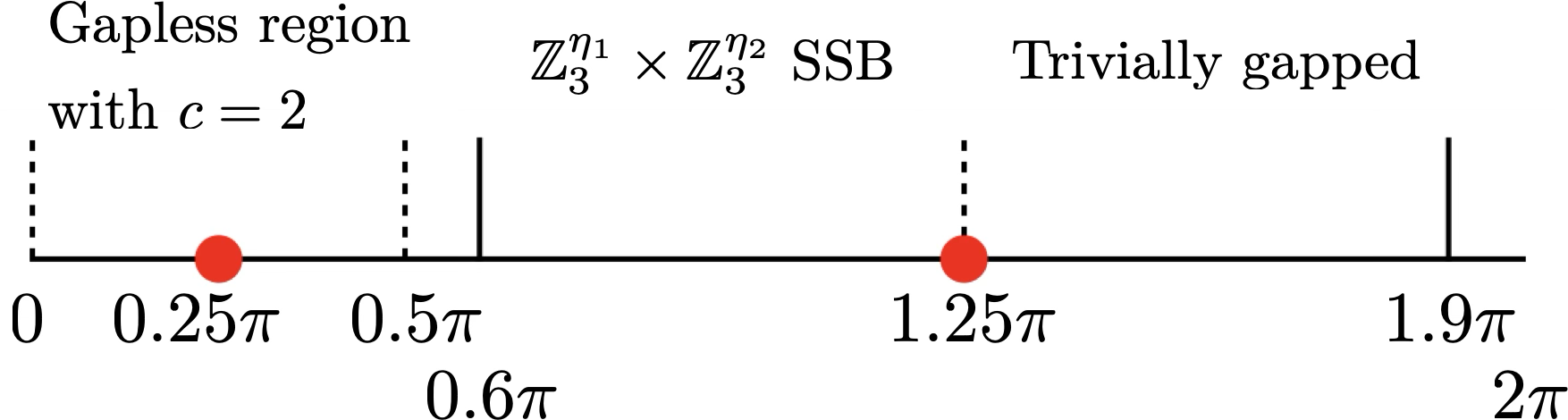}

\caption{The phase diagram of staggered-dipole-Ising model. The dashed lines are first-order phase transitions. The solid lines are continuous phase transitions and the red dots are self-dual points under RBPM.} 

\label{fig:phase diagram-2}
\end{figure}

Around $\theta=\pi/4$, the system is in the gapless phases with $c=2$ in the region $\theta\in (-0.1\pi,0)\cup(0,0.5\pi)\cup(0.5\pi,0.6\pi)$. These regions are separated by first-order phase transitions at $\theta=0,0.5\pi$, where the system is dominated by  $Z_{i-1}Z_{i}^2Z_{i+1}+\text{(H.c.)}$ and $X_i+\text{(H.c.)}$, respectively, leading to an exponentially large ground-state degeneracy. The gapless feature at the self-dual point $\theta= \pi/4$ is enforced by the anomaly of RBPM duality. Moreover, the system realizes a trivially gapped phase in the region $\theta\in(1.25\pi,1.9\pi)$, 
    dominated by $-X_i+\text{(H.c.)}$, and realizes a $\Z^{\eta_1}_3\times \Z^{\eta_2}_3$ SSB phase in the region $\theta\in (0.6\pi,1.25\pi)$ , dominated by  $-Z_{i-1}Z_{i}^2Z_{i+1}+\text{(H.c.)}$. The RBPM duality is a mapping between these two gapped phases, thus it determines the location of the phase transition at the self-dual point with $\theta=5\pi/4$, which is a first-order phase transition.

\prlsection{Summary and Discussion } In this work, we have presented the BPM framework for  Kramers-Wannier-type lattice dualities, whose linear-algebraic structure encodes (non)invertibility, symmetry realization, and anomaly data. 
%\hyo{In this work, we have reformulated \textcolor{red}{(LL: I think "reformulated" will make our paper looks trivial for referees..)} Kramers-Wannier-type lattice dualities as bilinear phase maps (BPMs), whose linear-algebraic structure encodes invertibility, symmetry realization, and anomaly data. 
Within this framework, we established an exact anomaly classification for a broad class of three-site RBPM %reflection$\times$BPM 
dualities: the duality is anomalous if and only if $-1$ is not a quadratic residue modulo $p$. 
This quadratic-residue criterion identifies the smallest anomalous prime as $p=3$, revealing lattice realizations of noninvertible symmetry constraints that are impossible in qubit systems.

Our results demonstrate that microscopic number-theoretic data -- encoded in the spectral structure of the BPM matrix over finite fields -- directly govern the existence of uniquely gapped symmetric phases. 
Beyond providing a structural understanding of lattice dualities, the BPM framework offers a systematic route to constructing and classifying generalized noninvertible symmetries in quantum spin systems.

The interplay between finite-field linear algebra and anomaly constraints uncovered here suggests broader connections between lattice dualities, generalized symmetries, and arithmetic structures, which we leave for future investigation.
 \\

\color{black}
\prlsection{Acknowledgments }
We thank Yunqin Zheng, Yuan Miao, Xiao Wang, and Weiguang Cao for the helpful discussions. LH.L. acknowledges the partial support from a Quantum SuperSEED grant (ICDS\_QS25\_029093) from the Institute for Computational and Data Sciences at the Pennsylvania State University and support from a startup
fund from the Pennsylvania State University (Zhen Bi).
H.Y. acknowledges the 2024 Toyota Riken Scholar Program from the Toyota Physical and Chemical Research Institute, the Overseas Research Support Grant from Yamada Science Foundation, and the  Grant-in-Aid for Research Activity Start-up from Japan Society for the Promotion of Science (Grant No. 24K22856).
The work of M.O. is partially supported by JSPS KAKENHI GrantS No. JP23K25791 and No. JP24H00946, and by
JST CREST Grant No. JPMJCR19T2.

\bibliography{xxx_ref.bib}
\clearpage
\appendix 
\onecolumngrid 
\color{black}

\setcounter{equation}{0}
\setcounter{figure}{0}
\setcounter{table}{0}
\makeatletter
\renewcommand{\theequation}{S\arabic{equation}}
\renewcommand{\thefigure}{S\arabic{figure}}
\renewcommand{\thetable}{S\arabic{table}}
\renewcommand{\bibnumfmt}[1]{[#1]}
\renewcommand{\citenumfont}[1]{#1}

 \begin{center}
	\Large{\textbf{Supplementary Materials for ``Generalized Kramers-Wannier Duality from Bilinear Phase Map''}}
\end{center}

Our Supplemental Material complements the discussions in the main text on a point-by-point basis.
For each main result presented in the paper, we provide explicit examples and detailed derivations to help readers explore the underlying physics in greater depth.
To serve this purpose, the Supplemental Material is organized in parallel with the structure of the main text, with each section corresponding to and expanding upon a specific part of the main discussion. The sections of the SM and their purposes are arranged as follows:

\begin{enumerate}
    \item Section \textbf{Example 1: A three-site BPM duality transformation}: \\
    First example for generalized BPM, via a three site-interaction model.
    \item Section \textbf{Generalized duality triangle of three-site BPM}: \\Details of the duality triangle from the three-site BPM and the corresponding spin chain models.
    \item Section \textbf{Example 2: A four-site BPM duality transformation}: \\ 
    Second example for models with generalized BPM, via a four site-interaction model.
    \item Section \textbf{Algebraic properties of the generalized BPM}: \\
    Derivation  of the kernels of generalized BPM and proof of its algebraic properties. 
    \item Section \textbf{Anomaly of three-site RBPM symmetry}: \\
    Proof of theorem 1 in the main text with several examples.
\end{enumerate}

\section{Example 1: A three-site BPM duality transformation}

\prlsection{The Model and BPM}
As the first example of generalized  BPM duality, we consider the following three-site Ising chain with length $L\in 3\Z$,
 \[\label{three site interaction}
\begin{split}
\mathcal H_{\text{3-Ising}}=-\sum^L_{i=1}(h X_i+Z_{i}Z_{i+1}Z_{i+2}).
 \end{split}
\]
When $h=\infty$, this system has a single paramagnetic ground state. When $h=0$, this system has SSB  ground states that we aim to understand. These two phases can be related by a generalized KW duality $\mathcal N_{\text{3-KW}}$, under PBC, given by the matrix  
\begin{equation}
\label{eq:3KWact1d}
    \vb*{A}_{\text{3-KW}}=\left(\begin{array}{cccccc}
    1&1&1&0& \cdots &0\\ 
    0&1&1&1& \cdots &0\\ 
    0&0&1&1& \cdots &0\\ 
    \cdots&\cdots&\cdots&\cdots&\cdots\\
    1&0&  \cdots  &0&1&1\\
    1&1&0&  \cdots &0&1\\
    \end{array}\right) .
\end{equation} 
The $\mathcal N_{\text{3-KW}}$ induces Pauli operator transformation 
\[
\begin{split}
 &\mathcal N_{\text{3-KW}}X_i=\hat{Z}_{i}\hat{Z}_{i+1}\hat{Z}_{i+2}\mathcal N_{\text{3-KW}}, \\ &\mathcal N_{\text{3-KW}}Z_{i-2}Z_{i-1}Z_{i}=\hat{X}_i\mathcal N_{\text{3-KW}}.
 \end{split}
\]
The properties of symmetry and SSB ground state degeneracy can be directly derived from the properties of $\vb*{A}_{\text{3-KW}}$. Its kernel is two dimensional,  generated by
\begin{equation}
\begin{split}
    &\vb*{b}^1 =  (1,1,0,1,1,0,\cdots), \\ &\vb*{b}^2 =  (0,1,1,0,1,1,\cdots),
    \end{split}
\end{equation}
which shows that the system has a $(\Z_2)^2$ symmetry:
\[
 U_{A} \ket{\{s_i\}} =   \ket{\{s_i + b^1_i \}}  ,\;
 U_{G} \ket{\{s_i\}} =   \ket{\{s_i + b^2_i \}}  .
\] 
In operator form, they are  written as 
\[ 
U_{A}=\prod^{L/3}_{i=1}X_{3i+1}X_{3i+2}, \quad
U_{G}=\prod^{L/3}_{i=1}X_{3i+2}X_{3i+3},
\]
 
The states can be organized into eigenstates of $U_{A/G}$ with eigenvalue $(-1)^{u_{A/G}}=\pm 1$, i.e., $u_{A/G}=0,1$. 
From the all spin-up state, this  $(\Z_2)^2$ symmetry can generate all ground states of SSB phases as $\ket{\vb*{0}}$, $\ket{\vb*{0}  + \vb*{b}^1 }$, $\ket{\vb*{0}  + \vb*{b}^2 }$, $\ket{\vb*{0}  + \vb*{b}^1+ \vb*{b}^2 }$: 
\[
\begin{split}
&\ket{\text{GS}}_1=\ket{\uparrow\uparrow\uparrow\cdots}, \quad \ket{\text{GS}}_2=\ket{\downarrow\downarrow\uparrow\cdots}
,\\ &\ket{\text{GS}}_3=\ket{\uparrow\downarrow\downarrow\cdots},\quad
\ket{\text{GS}}_4=\ket{\downarrow\uparrow\downarrow\cdots},
\end{split}
\]
and all these ground states are mapped to $|\rightarrow\rightarrow\cdots\rangle$ by $N_{\text{3-KW}}$ under PBC which indicates the non-invertible property.
Moreover, the dual model also has a $(\Z_2)^2$ symmetry generated by 
\[ 
\hat{U}_{A}=\prod^{L/3}_{i=1}\hat{X}_{3i+1}\hat{X}_{3i+2}, \quad
\hat{U}_{G}=\prod^{L/3}_{i=1}\hat{X}_{3i+2}\hat{X}_{3i+3}.
\]
Likewise, the dual Hilbert space can also be organized into four symmetry sectors labeled by
$(\hat{u}_A, \hat{u}_G) \in \{0,1\}^2$.  

Now, let us discuss the unitarity problem of BPM by introducing boundary spins $(\hat{t}_A,\hat{t}_G)\in \{0,1\}^2$ in $\{\hat{s}_i\}$-system, which corresponds to the untwisted/twisted boundary conditions of $(\Z_2)^2$ symmetry \cite{PhysRevB.106.045125,PhysRevB.106.075150,PhysRevB.106.075150,PhysRevB.106.224420,Yao:2023bnj}: 
\begin{align}
& \hat{s}_{L+3k+1}=\hat{s}_{3k+1}+\hat{t}_A, \hat{s}_{L+3k+2}=\hat{s}_{3k+2}+(\hat{t}_A+\hat{t}_G), \nonumber \\
& \hat{s}_{L+3k}=\hat{s}_{3k}+\hat{t}_G.
\end{align} 
Then we can modify the BPM as follows:
\[
\begin{split}
 &\mathcal N_{\text{3-KW}}\ket{\{s_{i}\}}\\&=\frac{1}{2^{\frac{L}{2}}} \sum_{\{\hat{s}_{i}\}}(-1)^{\sum_{j=1}^{L}\hat{s}_j (s_{j-2}+s_{j-1}+s_j)+\hat{t}_{G}s_{L}+\hat{t}_{A}s_{L-1}}\ket{\{\hat{s}_{i}\}}.
 \end{split}
\]
This modified BPM can distinguish four SSB ground states, satisfying that $s_{j-2}+s_{j-1}+s_{j}=0$. The BPM maps them to the same state
\[\frac{1}{2^{\frac{L}{2}}} \sum_{\{\hat{s}_{i}\}}(-1)^{ s_{L-1}\hat{t}_A +s_{L}\hat{t}_{G}}\ket{\{\hat{s}_{i}\}},
\]
that is the paramagnetic state  with a phase $(-1)^{\hat{t}_{A}s_{L-1}+\hat{t}_{G}s_{L}}$. When $\hat{t}_{A}=\hat{t}_{G}=0$, this phase is trivial and only linear combination $\sum^4_{i=1}\ket{\text{GS}}_i$ with $u_1=u_2=0$ survives. But when $\hat{t}_{A}=1$ and $\hat{t}_{G}=0$, two ground states with $s_{1}= 1$ will have additional $-1$ sign after mapping. Then only linear combination $(\ket{\text{GS}}_1+\ket{\text{GS}}_4-\ket{\text{GS}}_2-\ket{\text{GS}}_3)$ survives under BPM duality. This combination has symmetry charge $u_A=u_G=1$. On the other hand, when $\hat{t}_{A}=0$ and $\hat{t}_{G}=1$, two ground states with $s_{L}= 1$ will have additional $-1$ sign after mapping. Only linear combination $(\ket{\text{GS}}_1-\ket{\text{GS}}_4+\ket{\text{GS}}_2-\ket{\text{GS}}_3)$ survives under BPM, which has symmetry charge $u_A=0,u_G=1$. 
Moreover, in the SM, we also show $\mathcal N_{\text{3-KW}}$ is anomaly-free, allowing an SPT phase to preserve duality-symmetry and find a generalized Kennedy-Tasaki (KT) duality \cite{1992CMaPh.147..431K,PhysRevB.45.304,oshikawa1992hidden,PhysRevB.46.3486,PhysRevB.107.125158,PhysRevB.78.094404,PhysRevB.88.085114,PhysRevB.83.104411,Devakul:2018fhz,PhysRevB.108.214429,li2023intrinsicallypurely,Bhardwaj:2023bbf,ParayilMana:2024txy} between this SPT phase and the   SSB phase.\\

\prlsection{Symmetry-twisting mapping of three-site BPM duality}
We now derive the symmetry-twist sectors of BPM $\mathcal N_{\text{3-KW}}$. Similar to the boundary spins in $\{\hat{s}_i\}$-system, we also introduce boundary spins $(t_1,t_2)\in \{0,1\}^2$ in $\{s_i\}$-system:
\[
\begin{split}
&s_{L+3k+1}=s_{3k+1}+t_A, \quad s_{L+3k+2}=s_{3k+2}+(t_A+t_G),\quad s_{L+3k}=s_{3k}+t_G.
\end{split}
\]
Then we find a consistent modified expression of $\mathcal N_{\text{3-KW}}$ in $\{s_i\}$ and $\{\hat{s}_i\}$ systems:
\[
\begin{split}
 \mathcal N_{\text{3-KW}}\ket{\{s_{i}\}}&=\frac{1}{2^{\frac{L}{2}}} \sum_{\{\hat{s}_{i}\}}(-1)^{\sum_{j=1}^{L}s_j (\hat{s}_{j}+\hat{s}_{j+1}+\hat{s}_{j+2})+t_A\hat{s}_{1}+t_G\hat{s}_{2}}\ket{\{\hat{s}_{i}\}}\\&=\frac{1}{2^{\frac{L}{2}}} \sum_{\{\hat{s}_{i}\}}(-1)^{\sum_{j=1}^{L}\hat{s}_j (s_{j-2}+s_{j-1}+s_j)+\hat{t}_{G}s_{L}+\hat{t}_{A}s_{L-1}}\ket{\{\hat{s}_{i}\}}.
 \end{split}
\]
From this formula, it is straightforward to check the symmetry-twist mapping:
\[
\begin{split}\label{eq:sym-twist map-3 site}
[(\hat{u}_{A},\hat{t}_{A}),(\hat{u}_{G},\hat{t}_{G})]=[(t_A+t_G,u_A),(t_G,u_A+u_G)].
\end{split}
\]
Let us first acts $\hat{U}_A\times \mathcal N_{\text{3-KW}}$ and $\hat{U}_G\times \mathcal N_{\text{3-KW}}$ on the state $\ket{\{s_{i}\}}$:
\[
\begin{split}
 &\hat{U}_A \mathcal N_{\text{3-KW}}\ket{\{s_{i}\}}=(-1)^{t_A+t_G} \mathcal N_{\text{3-KW}}\ket{\{s_{i}\}},
 \\&\hat{U}_G \mathcal N_{\text{3-KW}}\ket{\{s_{i}\}}=(-1)^{t_G} \mathcal N_{\text{3-KW}}\ket{\{s_{i}\}}.
 \end{split}
\]
This holds for any state $N_{\text{3-KW}}|\psi\rangle$, where $|\psi\rangle$ is a general state in $\{s_i\}$ system $\ket{\psi} = \sum_{\{s_i\}} \psi_{\{s_i\}} \ket{\{s_i\}}$. Thus any state obtained by acting $N_{\text{3-KW}}$ must be eigenstate of $\hat{U}_A$ and $\hat{U}_G$
with eigenvalue $(\hat{u}_A,\hat{u}_G)=(t_A+t_G,t_G)$.

Next, we continue to consider $ \mathcal N_{\text{3-KW}}\times U_1$ and $ \mathcal N_{\text{3-KW}}\times U_G$:
\[
\begin{split}
 & \mathcal N_{\text{3-KW}}U_A\ket{\{s_{i}\}}=(-1)^{\hat{t}_A} \mathcal N_{\text{3-KW}}\ket{\{s_{i}\}},
 \\& \mathcal N_{\text{3-KW}}U_G\ket{\{s_{i}\}}=(-1)^{\hat{t}_A+\hat{t}_G} \mathcal N_{\text{3-KW}}\ket{\{s_{i}\}}.
 \end{split}
\]
Similarly, this is valid for general state $|\psi\rangle$. In particular, we can consider an eigenstate $\ket{\Psi}$ of $(U_A,U_G)$ with eigenvalue $(u_A,u_G)$. Thus we have
\[
\begin{split}
&\mathcal N_{\text{3-KW}}U_A \ket{\Psi}=(-1)^{\hat{t}_A}N_{\text{3-KW}}\ket{\Psi}=N_{\text{3-KW}}(-1)^{u_A}\ket{\Psi},\\
&\mathcal N_{\text{3-KW}}U_G \ket{\Psi}=(-1)^{\hat{t}_A+\hat{t}_G}N_{\text{3-KW}}\ket{\Psi}=N_{\text{3-KW}}(-1)^{u_G}\ket{\Psi},
\end{split}
\]
namely,
\[
(u_A,u_G)=(\hat{t}_A,\hat{t}_A+\hat{t}_G).
\]
Then it follows that $(\hat{t}_A,\hat{t}_G)=(u_A,u_A+u_G)$.

\section{Generalized duality triangle of three-site BPM} 

\prlsection{Self-dual point} The self-dual point $|h|=1$ is expected to be the phase transition point between $\mathbb{Z}^2_2$ SSB phase and the trivially gapped phase \cite{PhysRevB.108.214430}, where the duality transformation becomes a non-invertible symmetry. However, unlike the usual KW duality symmetry which is anomalous, the BPM \eqref{eq:3KWact1d} is anomaly free, namely it allows the self-dual uniquely gapped phases, e.g., the $\Z^A_2\times \Z^G_2$ SPT phase. 
A solvable Hamiltonian is given by \[\label{SPT Hal}
\mathcal H_{\text{SPT}}=-\sum^{L}_{i=1}a_i, \quad a_i=(-1)^i Z_{i-1}Y_i Z_{i+1}.
\] 
Such SPT Hamiltonian can be constructed by decorated domain wall (DW) method \cite{chen2014symmetry,Wang:2021nrp,Li:2022jbf}. As one can check, the product of two nearest neighbored terms is $Z_{i-1}X_iX_{i+1}Z_{i+2}$, which comes from decorating the domain wall term $Z_{i-1} Z_{i+2}$ with charge operator $X_iX_{i+1}$. For example, if we assume $i=1 \text{(mod 3)}$, the $Z_{i-1} Z_{i+2}$ is a domain wall term of $U_G$ and the charge operator $X_iX_{i+1}$ is associated with $U_A$ \footnote{More precisely, the ground state is an eigenstate of $Z_{i-1}X_iX_{i+1}Z_{i+2}$ with eigenvalue 1 and thus has the SPT feature. The reason why not choosing $H=-\sum^L_{i=1} Z_{i-1}X_iX_{i+1}Z_{i+2}$ is that this Hamiltonian has an emergent symmetry $\prod_i X_i$ and is in the corresponding SSB phase.}. This construction can be implemented by a unitary transformation $U_{\text{3-DW}}$, which can map the SPT Hamiltonian to trivially gapped Hamiltonian:
\[
U^{\dagger}_{\text{3-DW}}\mathcal H_{\text{SPT}}U_{\text{3-DW}}=-\sum^{L}_{i=1}X_i\equiv \mathcal  H_{\text{triv}}.
\]
where 
\[U_{\text{3-DW}}=\prod^{L}_{i=1}\exp(-\frac{\pi i}{4}(-1)^i Z_i)\prod^{L}_{i=1}\exp[\frac{\pi i}{4}(1-Z_i)(1-Z_{i+1})] T,
\]
and $T$ is one-site lattice translation. The dual Hamiltonian has a unique ground state, thus the Hamiltonian \eqref{SPT Hal}
also has a unique gapped ground state. \\ 

\prlsection{String order parameters, ground state charge under twisted boundary condition and edge modes}
In this section, we will detect the SPT order by different methods. The first method is the string order parameter:
\[
\begin{split}
\langle S_{U_A}\rangle &=(-1)^{m-n+1}\langle Z_{3n}\prod^m_{k=n}(X_{3k+1}X_{3k+2})Z_{3m+3}\rangle=\langle\prod^m_{k=n}a_{3k+1}a_{3k+2}\rangle=1,\\\langle S_{U_G}\rangle&=(-1)^{m-n+1}\langle Z_{3n-2}\prod^m_{k=n}(X_{3k-1}X_{3k})Z_{3m+1}\rangle=\langle\prod^m_{k=n}a_{3k-1}a_{3k}\rangle=1.
\end{split}
\]
The string order parameter 
$S_{U_A}$ ($S_{U_G}$) is obtained by dressing the string operator of $U_A$ ($U_G$) symmetry with charged operator of $U_G$ ($U_A$) symmetry at endpoints, which is consistent with decorated domain wall construction.

The second way to probe the SPT order is ground state charge under
twisted boundary conditions on the closed chains. For simplicity, we assume $L\in 6\mathbb{Z}$. Let us first twist the boundary condition using the $\Z_2^{A}$ symmetry (labeled by $\Z_2^{A}$-TBC), and measure the $\Z_2^{G}$ charge of the ground state. Twisting the boundary condition by $\Z_2^{A}$ means imposing a domain wall between sites $L-1$ and $1$ by changing the sign of the term $Z_{L-1}Y_L Z_{1}$. The SPT Hamiltonian \eqref{SPT Hal} becomes 
\[
\mathcal H^{\Z_2^A}_{\text{SPT}}=\sum^{L-1}_{i=1}a_i-a_L.
\]
We note that the twisted and untwisted SPT Hamiltonian are related by a unitary transformation $H_{\text{SPT}}^{\Z_2^A}= Z_{L}H_{\text{SPT}}Z_L$. Denote the ground state under PBC as $\ket{\text{GS}}$, and that under $\Z_2^A$-TBC as $\ket{\text{GS}}_{\text{tw}}^{\Z_2^A}$. We have
\begin{eqnarray}
\ket{\text{GS}}_{\text{tw}}^{\Z_2^A}= Z_L \ket{\text{GS}}.
\end{eqnarray}
It follows that 
\begin{eqnarray}
U_G \ket{\text{GS}}_{\text{tw}}^{\Z_2^A} = U_G Z_L  \ket{\text{GS}} =-Z_L \ket{\text{GS}}= -\ket{\text{GS}}_{\text{tw}}^{\Z_2^A}
\end{eqnarray}
which shows that $\ket{\text{GS}}_{\text{tw}}^{\Z_2^A}$ has  $\Z_2^G$ charge 1. Here we used the fact that the ground state under PBC is neutral under $\Z_2^G$. 
We can alternatively twist the boundary condition using $\Z_2^G$ symmetry (labeled by $\Z_2^G$-TBC), and measure the $\Z_2^A$ charge of the ground state. By the same method, one can show that the ground state $\ket{\text{GS}}_{\text{tw}}^{\Z_2^G}$ has odd $\Z_2^A$ charge. 

At last, we will derive how symmetry fractionalizes on edge modes. Let us place the spin system on an open chain with $1 \le i \le L$ and choose the OBC such that only the interactions completely supported on the chain are kept.
The Hamiltonian is
\begin{eqnarray}
\mathcal H_{\text{SPT}}^{\text{OBC}}=- \sum^{L-1}_{j=2} a_j.
\end{eqnarray}
There are two boundary terms on each edge: $Z_1$, $X_1 Z_2$, $Z_L$, $Z_{L-1} X_L$. All the operators commute with bulk Hamiltonian and the anticommutation relation of two terms on each edge gives rise to 2-fold degenerate subspace. 

To see the symmetry fractionalization, we note that for ground states 
\[
\begin{split}
&\prod^{L/3-1}_{k=1}a_{3k+1}a_{3k+2}=-Z_{3}\prod^{L/3-1}_{k=1}(X_{3k+1}X_{3k+2})Z_{L}=1,\\
&\prod^{L/3-1}_{k=1}a_{3k-1}a_{3k}=-Z_{1}\prod^{L/3-1}_{k=1}(X_{3k-1}X_{3k})Z_{L-2}=1.
\end{split}
\]
This implies the symmetry operator fractionalizes as $U_{A/G}=-\mathcal{L}_{A/G}\mathcal{R}_{A/G}$
where
\[
\begin{split}
 &\mathcal{L}_{A}=X_1X_2Z_3, \mathcal{R}_{A}=Z_L, \\ &\mathcal{L}_{A}=Z_1, \mathcal{R}_{G}=Z_{L-2}X_{L-1}X_L.
\end{split}
\]
On each edge, the projective representation of $\mathcal{L}$ and $\mathcal{R}$ gives rise to the edge modes. This projective representation and the resulting edge modes are robust as long as the bulk gap is not closed. Here, we remark that the discussion above shows that the above model  realizes a nontrivial $\mathbb{Z}^A_2 \times \mathbb{Z}^G_2$ SPT phase. In fact, this model possesses a larger non-invertible symmetry, due to also being invariant under $\mathcal{N}_{\text{3-KW}}$. By contrast, the trivial $\mathbb{Z}^A_2 \times \mathbb{Z}^G_2$ SPT phase described by the Hamiltonian \eqref{three site interaction} with $h=\infty$ is not invariant under $\mathcal{N}_{\text{3-KW}}$. This situation closely parallels the recent result in Ref.~\cite{PhysRevLett.133.116601}, where the cluster state, known as the nontrivial $\mathbb{Z}_2 \times \mathbb{Z}_2$ SPT, was also found to exhibit a larger non-invertible $\text{Rep}(D_8)$ symmetry. We leave the classification of the SPTs enriched by $\mathcal{N}_{\text{3-KW}}$ to future work.

At last, we also comment that when $L\notin 3$, the Hamiltonian \eqref{SPT Hal} does not respect $(\Z_2)^2$ symmetry. This is because BPM under PBC only has the trivial kernel $(0,0,0,\cdots)$ and is not associated with $U_A$ and $U_G$ in this case.  Thus the $\mathcal{N}_{3-\text{KW}}$ is a unitary transformation and the Hamiltonian \eqref{three site interaction} with $h=0$ has a unique ground state.\\

\begin{figure}[htpb] 
\centering 
\includegraphics[width=0.5\textwidth]{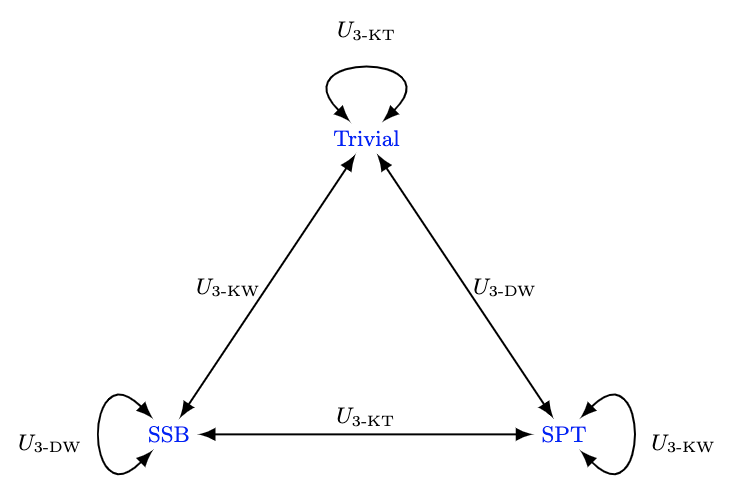} 
\caption{ Three gapped phases with $\Z_2\times \Z_2$ symmetry and the dualities between them. } 
\label{fig:duality web} 
\end{figure}

\begin{figure}[htpb] 
\centering 
\includegraphics[width=0.5\textwidth]{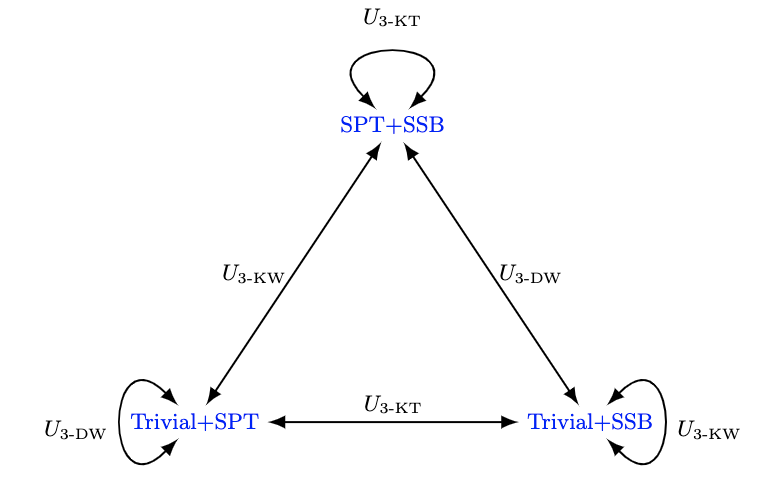} 
\caption{Three phase transitions between two different gapped phases with $\Z_2\times \Z_2$ symmetry and the dualities between them. } 
\label{fig:duality web-2} 
\end{figure}

\prlsection{The generalized Kennedy-Tasaki transformation }
Now, since there are SPT, trivially gapped  and SSB phases, we can construct a web of duality connecting them, which is summarized in Fig.~\ref{fig:duality web}.
Here $U_{\text{3-KT}}$ is a generalized Kennedy-Tasaki transformation \cite{PhysRevB.108.214429,li2023intrinsicallypurely}:
\[
U_{\text{3-KT}}=U_{\text{3-DW}}U_{\text{3-KW}}U^{\dagger}_{\text{3-DW}}.
\]
Moreover, such a web of duality can also connect phase transitions between two different gapped phases, as shown in Fig.~\ref{fig:duality web-2}. Identifying the nature of these phase transitions by numerical calculation is left for the future.

Similar to the reference \cite{PhysRevB.108.214429},  this generalized Kennedy-Tasaki transformation is constructed as follows:
\[
U_{\text{3-KT}}=U_{\text{3-DW}}U_{\text{3-KW}}U^{\dagger}_{\text{3-DW}}.
\]
 It is straightforward to derive the fusion rule of this KT transformation:
\[
\begin{split}
U_{\text{3-KT}}U_{\text{3-KT}}&=U_{\text{3-DW}}U_{\text{3-KW}}U_{\text{3-KW}}U^{\dagger}_{\text{3-DW}}\\&=U_{\text{3-DW}}(1+U_A)(1+U_G)T^2 U^{\dagger}_{\text{3-DW}}\\&=(1+U_A)(1+U_G)T^2
\end{split}
\]
where we use the fact that $U_{\text{3-DW}}$ commutes with two-site translation and $U_A$ and $U_G$ operators.

\section{Example 2: A four-site BPM duality transformation}
\prlsection{BPM duality under PBC and fusion rules}
We discuss the the second example of BPM duality which is related to the following four-site Ising chain with $L\in 4\mathbb{Z}$:
    \[
\begin{split}
\mathcal H_{\text{4-Ising}}=-\sum^L_{i=1}(h X_i+Z_{i}Z_{i+1}Z_{i+2}Z_{i+3}).
 \end{split}
\]
This system has a unique ground state with all $X_i=1$ when $h=\infty$, while it is in the SSB phase when $h=0$. To understand the duality between these two phases, we can construct a
generalized KW duality under PBC:
\[
\begin{split}
 \mathcal N_{\text{4-KW}}\ket{\{s_{i}\}}&=\frac{1}{2^{\frac{L}{2}}} \sum_{\{\hat{s}_{i}\}}(-1)^{\sum_{j=1}^{L}s_j (\hat{s}_{j}+\hat{s}_{j+1}+\hat{s}_{j+2}+\hat{s}_{j+3})}\ket{\{\hat{s}_{i}\}}\\&=\frac{1}{2^{\frac{L}{2}}} \sum_{\{\hat{s}_{i}\}}(-1)^{\sum_{j=1}^{L}\hat{s}_j (s_{j-3}+s_{j-2}+s_{j-1}+s_j)}\ket{\{\hat{s}_{i}\}} , 
 \end{split}
\]
whose BPM matrix is :
\begin{equation} 
    \vb*{A}_{\text{4-KW}}=\left(\begin{array}{ccccccc}
    1&1&1&1&0& \cdots &0\\ 
    0&1&1&1&1& \cdots &0\\ 
    0&0&1&1&1& \cdots &0\\ 
    \cdots&\cdots&\cdots&\cdots&\cdots\\
    1&1&0&  \cdots  &0&1&1\\
    1&1&1&0&  \cdots &0&1\\
    \end{array}\right) .
\end{equation} 
Similarly, the properties of symmetry and SSB ground state degeneracy can be derived from the kernels of $\vb*{A}^T_{\text{4-KW}}$:  
\begin{align}
    \vb*{b}^1 =  (1,0,1,0,1,0,1,0,\cdots), \quad
    \vb*{b}^2 =  (0,1,0,1,0,1,0,1,\cdots) ,\quad
    \vb*{b}^3 =  (1,1,0,0,1,1,0,0,\cdots)  .
\end{align}
This shows the system has a $(\mathbb{Z}_2)^3$ symmetry:
\begin{align}
 &U_{o} \ket{\{s_i\}} =   \ket{\{s_i + b^1_i \}},  \quad
 U_{e} \ket{\{s_i\}} =   \ket{\{s_i  + b^2_i\}}  ,\quad
 U_{1} \ket{\{s_i\}} =   \ket{\{s_i +   b^3_i\}}  .
\end{align}
In operator form, they are given by:
 \[
\begin{split}
&U_{e}=\prod^{L/2}_{i=1} X_{2i},\quad U_{o}=\prod^{L/2}_{i=1} X_{2i+1},\quad U_{1}=\prod^{L/4}_{i=1}X_{4i+1}X_{4i+2}. 
 \end{split}
\]
Thus all states can be organized into eigenstates of $U_{e/o/1}$ with eigenvalue $(-1)^{u_{e/o/1}}=\pm 1$, i.e. $u_{e/o/1}=0,1$.

The eight ground states of the SSB phase can be generated by these symmetry operators from the all spin-up state:
\[
\begin{split}
&\ket{\text{GS}}_1=\ket{\uparrow\uparrow\uparrow\uparrow\cdots}, \quad \ket{\text{GS}}_2=\ket{\uparrow\downarrow\uparrow\downarrow\cdots}, \\
&\ket{\text{GS}}_3=\ket{\downarrow\uparrow\downarrow\uparrow\cdots},\quad
\ket{\text{GS}}_4=\ket{\downarrow\downarrow\downarrow\downarrow\cdots},\\& \ket{\text{GS}}_5=\ket{\downarrow\downarrow\uparrow\uparrow\cdots},\quad
\ket{\text{GS}}_6=\ket{\downarrow\uparrow\uparrow\downarrow\cdots},\\
&\ket{\text{GS}}_7=\ket{\uparrow\downarrow\downarrow\uparrow\cdots},\quad \ket{\text{GS}}_8=\ket{\uparrow\uparrow\downarrow\downarrow\cdots}.
\end{split}
\]

This BPM duality induces the transformation of Pauli operators   \[
\begin{split}
 &\mathcal N_{\text{4-KW}}X_i=\hat{Z}_{i}\hat{Z}_{i+1}\hat{Z}_{i+2}\hat{Z}_{i+3}\mathcal N_{\text{4-KW}}, \\
& \mathcal N_{\text{4-KW}}Z_{i-3}Z_{i-2}Z_{i-1}Z_{i}=\hat{X}_i\mathcal N_{\text{4-KW}},
 \end{split}
\]
and thus exchanges transverse field term and four-site Ising term. Likewise, the dual Hilbert space can also be organized into four symmetry sectors labeled by
$(\hat{u}_o, \hat{u}_e, \hat{u}_1) \in \{0,1\}^3$. 

We can also determine fusion rules by acting with the products of $\hat{U}_{e/o/1}\times \mathcal{N}_{\text{4-KW}}$, $\mathcal{N}_{\text{4-KW}}\times U_{e/o/1}$ and $\mathcal{N}_{\text{4-KW}}\times\mathcal{N}_{\text{4-KW}}$ on a general state:
\begin{equation} 
\begin{split}
       &\mathcal{N}_{\text{4-KW}}\times U_{e/o/1}=\mathcal{N}_{\text{4-KW}},\quad\hat{U}_{e/o/1}\times\mathcal{N}_{\text{4-KW}}=\mathcal{N}_{\text{4-KW}}, \\& \mathcal{N}_{\text{4-KW}}\times \mathcal{N}_{\text{4-KW}}=(1+U_{e})(1+U_{o})(1+U_{1})T^3\, . 
\end{split}
\end{equation}  

\prlsection{Unitarity problem and symmetry-twist transformation}
To solve this unitarity problem in this case, we need to add three additional boundary spins $t_e$, $t_o$ and $t_1$ in $\{s_i\}$-systems and another three spins $\hat{t}_e$, $\hat{t}_o$ and $\hat{t}_1$ in $\{\hat{s}_i\}$-systems, i.e. the untwisted/twisted boundary condition of $(\Z_2)^3$ symmetry: 
\begin{equation}
\begin{split}
  & s_{L+4k+1}=s_{4k+1}+(t_o+t_1),   s_{L+4k+2}=s_{4k+2}+(t_e+t_1),\\ & s_{L+4k+3}=s_{4k+3}+t_o,\quad s_{L+4k}=s_{4k}+t_e.\\ 
   & \hat{s}_{L+4k+1}=\hat{s}_{4k+1}+(\hat{t}_o+\hat{t}_1),   \hat{s}_{L+4k+2}=\hat{s}_{4k+2}+(t_e+t_1),\\ & \hat{s}_{L+4k+3}=\hat{s}_{4k+3}+\hat{t}_o,\quad \hat{s}_{L+4k}=\hat{s}_{4k}+\hat{t}_e.
  \end{split}
\end{equation}

We also find a consistent modification of BPM: 

\[
\begin{split}
 &\mathcal N_{\text{4-KW}}\ket{\{s_{i}\}}=\frac{1}{2^{\frac{L}{2}}} \sum_{\{\hat{s}_{i}\}}(-1)^{\sum_{j=1}^{L}s_j (\hat{s}_{j}+\hat{s}_{j+1}+\hat{s}_{j+2}+\hat{s}_{j+3})+t_e(\hat{s}_{2}+\hat{s}_{3})+t_o(\hat{s}_{1}+\hat{s}_{2})+t_1 \hat{s}_{1}}\ket{\{\hat{s}_{i}\}}\\
 &=\frac{1}{2^{\frac{L}{2}}} \sum_{\{\hat{s}_{i}\}}(-1)^{\sum_{j=1}^{L}\hat{s}_j (s_{j-3}+s_{j-2}+s_{j-1}+s_j)+\hat{t}_{o}(s_{L-1}+s_{L-2})+\hat{t}_{e}(s_{L}+s_{L-1})+\hat{t}_{1}s_{L-2}}\ket{\{\hat{s}_{i}\}}.
 \end{split}\]
By a similar method, one can find the symmetry-twist mapping from this formula:
\[
\begin{split}\label{eq:sym-twist map}
&[(\hat{u}_{o},\hat{t}_{o}),(\hat{u}_{e},\hat{t}_{e}), (\hat{u}_{1},\hat{t}_{1})]\\
=&[(t_o+t_e+t_1,u_e+u_o+u_1),(t_o+t_e,u_e+u_1), (t_1+t_e,u_o+u_e)].
\end{split}
\]
We can apply this modified BPM to fix the unitarity problem for SSB ground states, which satisfies that $s_{j-3}+s_{j-2}+s_{j-1}+s_{j}=0$. The BPM maps them to the paramagnetic state with $\hat{X}_i=1$:
\[\frac{1}{2^{\frac{L}{2}}} (-1)^{\hat{t}_{o}(s_{L-1}+s_{L-2})+\hat{t}_{e}(s_{L}+s_{L-1})+\hat{t}_{1}s_{L-2}}\sum_{\{\hat{s}_{i}\}}\ket{\{\hat{s}_{i}\}}.
\]
When $\hat{t}_{o}=\hat{t}_{e}=\hat{t}_{1}=0$, the phase is trivial and only linear combination $\sum^8_{i=1}\ket{\text{GS}}_i$ with all $u=0$ survives. But when $\hat{t}_{o}=1$ and $\hat{t}_{e}=\hat{t}_{1}=0$, four ground states with $s_{L-1}+s_{L-2}= 1$ will be mapped with additional $-1$ sign. Then only linear combination $(\ket{\text{GS}}_1+\ket{\text{GS}}_4+\ket{\text{GS}}_6+\ket{\text{GS}}_7-\ket{\text{GS}}_2-\ket{\text{GS}}_3-\ket{\text{GS}}_5-\ket{\text{GS}}_8)$ survives. This combination has symmetry charge $u_o=u_e=u_1=1$, which is the solution of Eq.~\eqref{eq:sym-twist map}. It is straightforward to check other cases and linear combinations of SSB ground states with different symmetry eigenvalues will be mapped to the paramagnetic state under different boundary conditions, which satisfies the rule of symmetry-twist mapping \eqref{eq:sym-twist map}.\\

\prlsection{Anomaly of four-site BPM duality symmetry }
On the self-dual point $h=1$, the model $\mathcal H_{\text{4-Ising}}$ is at a first-order phase transition between SSB phase and the trivial phase \cite{PhysRevB.29.2404,PhysRevE.90.032101}. The BPM duality $\mathcal N_{\text{4-KW}}$ also becomes an emergent non-invertible symmetry for the self-dual theory. Such an emergent symmetry is anomalous in the sense that it cannot allow a symmetric uniquely gapped phase under any symmetric perturbations and hence self-dual theories must be always at first-order or second-order first phase transitions.

To prove the anomaly of $\mathcal N_{\text{4-KW}}$, let us first show this duality operator can be decomposed as the product under PBC: $\mathcal{N}_{\text{4-KW}}=\frac{1}{2}\mathcal{N}_{\text{KW'}}\times (\mathcal{N}_{\text{KW}})^{\dagger}\times \mathcal{N}_{\text{KW'}}$.
Here the $\mathcal{N}_{\text{KW}}$ is the usual KW transformation and $\mathcal{N}_{\text{KW'}}$ is the combination of two KW transformations acting on even and odd sites:
\[
\begin{split}
   &\mathcal N_{\text{KW'}}\ket{\{s_{i}\}}= \frac{1}{2^{\frac{L}{2}}} \sum_{\{\hat{s}_{i}\}}(-1)^{\sum_{j,k=1}^{L}(s_{j-2}+s_j) \hat{s}_{k}}\ket{\{\hat{s}_{i} \}}.  
\end{split}
\]
We can directly check this result
\begin{equation}\label{4-kW relation}
\begin{split}
&\mathcal{N}_{\text{KW'}}\times (\mathcal{N}_{\text{KW}})^{\dagger}\times \mathcal{N}_{\text{KW'}}\ket{\{s_{i}\}}\\
=& \frac{1}{2^{\frac{3L}{2}}} \sum_{\{s'_i,s''_i,\hat{s}_{i}\}}  (-1)^{\sum^L_{j=1}s'_j(s_{j-2}+s_{j})+s'_j(s''_{j-1}+s''_{j})+\hat{s}_{j}(s''_{j-2}+s''_j)}\ket{\{\hat{s}_{i}\}}\\
=& \frac{1}{2^{\frac{L}{2}}} \sum_{\{s''_i,\hat{s}_{i}\}} \delta(s_{j-2}+s_j+s''_{j-1}+s''_{j}) (-1)^{\sum^L_{j=1}\hat{s}_{j}(s''_{j-2}+s''_j)}\ket{\{\hat{s}_{i}\}}\\
=& \frac{1}{2^{\frac{L}{2}}} \sum_{\{s''_i=s_i+s_{i-1}+0/1,\hat{s}_{i}\}}  (-1)^{\sum^L_{j=1}\hat{s}_{j}(s''_{j-2}+s''_j)}\ket{\{\hat{s}_{i}\}}\\
=& \frac{2}{2^{\frac{L}{2}}} \sum_{\{\hat{s}_{i}\}} (-1)^{\sum^L_{j=1}\hat{s}_{j}(s_{j-3}+s_{j-2}+s_{j-1}+s_j)}\ket{\{\hat{s}_{i}\}}
\\
=& 2\mathcal{N}_{\text{4-KW}}\ket{\{\hat{s}_{i}\}}.
\end{split}
\end{equation}

Now, let us prove the anomaly by the contraction method. We first assume a uniquely gapped system is self-dual under PBC and its ground state $|\psi\rangle$ should be short-range entangled (SRE). Due to symmetry-twist mapping, $|\psi\rangle$ should be even under each $\Z_2$ symmetry. If we focus on the $\Z^e_2\times \Z^o_2$ symmetry, the possible uniquely gapped phase can only be the $\Z^e_2\times \Z^o_2$ SPT phase, since trivially gapped phase is mapped to an SSB phase under this four-site BPM. Then we can perform $(\mathcal{N}_{\text{KW'}})^{\dagger}$ or $\mathcal{N}_{\text{KW'}}$ which both keep the SPT phase invariant \cite{PhysRevB.108.214429}. Thus $\mathcal{N}_{\text{KW'}}^{\dagger}|\psi\rangle$ and $\mathcal{N}_{\text{KW'}}|\psi\rangle$ are still ground states of the $\Z^e_2\times \Z^o_2$ SPT systems and thus SRE.  On the other hand, due to \eqref{4-kW relation}, we have
\[
\begin{split}
\langle \psi|\frac{1}{2}\mathcal{N}_{\text{KW'}}\times \frac{1}{2\sqrt{2}}\mathcal{N}_{\text{KW}}^{\dagger} \mathcal{N}_{\text{KW'}}|\psi\rangle
=\langle \psi|\frac{1}{2\sqrt{2}}\mathcal{N}_{\text{4-KW}}|\psi\rangle=e^{i\theta}
\end{split}
\]
where we multiply the normalized coefficient $\frac{1}{2\sqrt{2}}$. That is \[
\frac{1}{2\sqrt{2}}\mathcal{N}_{\text{KW}}^{\dagger} \mathcal{N}_{\text{KW'}}|\psi\rangle=e^{i\theta}\frac{1}{2}\mathcal{N}_{\text{KW'}}^{\dagger} |\psi\rangle.
\] 
 However, the $\mathcal{N}^{\dagger}_{\text{KW}}$ maps the $\Z^e_2\times \Z^o_2$ SPT phase to an SSB phase of global spin flip $U_e U_o$. Thus $\frac{1}{2\sqrt{2}}\mathcal{N}^{\dagger}_{\text{KW}}\mathcal{N}_{\text{KW'}}|\psi\rangle$ is a cat state of SSB phase with even charge of $U_e U_o$ which is not SRE and that finishes our proof.\\

\section{Algebraic properties of the generlized BPM}

Let us consider the spin chain of $\mathbb{Z}_p$ qudits, where $p$ is a prime number (our analysis can be extended to general $p$ --- it only requires more tedious modular arithmetic, but nothing new conceptually.)
We consider the translational symmetric Hamiltonian on a chain of qudits of $k+1$-site interactions,
\begin{equation}
    \mathcal{H} = \sum_i Z_i^{n_0}Z_{i+1}^{n_1}\cdots Z_{i+k}^{n_k} + X_i +\text{h.c.}
\end{equation}
Here, $n_j, \ j= 0, \dots, k$ are integers $\mod p$, i.e., numbers in finite field $\mathbb{F}_p$ (here, field is the concept in algebra, which is numbers with well defined addition, subtraction, multiplication and division). 
By  $k+1$-site interaction, we mean both $n_0$ and $n_k$ must be non-zero --- otherwise, the interaction is effectively of shorter distance. However, other $n_j$'s can be zero. 

Our analysis of BPM can be used to determine the symmetries of this model. 
The line in BPM matrix $\vb*{A}^{\text{T}}$ corresponding to the term labeled $i$ is
\begin{equation}
    0 ,0,\cdots, n_0, n_1, \cdots , n_k, 0 , 0 ,\dots.
\end{equation}
More precisely, we have $A_{ji}=n_{j-i}$ when $i\le j\le i+k$ while $A_{ji}=0$ when $i>j$ or $i+k<j$.
So the kernel $\vb*{b}$ of the BPM, a vector defined on field $\mathbb{F}_p$, and tells us the symmetry of the model, must satisfy
\begin{equation}
\label{eqn.recur}
    \sum_{m = 0}^{k}n_m b_{i+m}  = 0
\end{equation}
and this constraint must be satisfied for all $i$'s. 

Let us consider a chain with open boundary condition on the left side, starting at $i=1$. 
We will later discuss the  open, periodical boundary conditions and thermodynamics limit. 
Equation~\eqref{eqn.recur} defines a \textit{recurrence relation} or \textit{partial difference equation} imposed on $\vb*{b}$. That is, the $\mathbb{F}_p$ number $b_{i+k}$ is uniquely determined by the previous $k$ numbers $b_i$ to $b_{i+k-1}$.

Let us consider the vector with $k$ elements, 
\begin{eqnarray}
    \vb*{v}_{(i)}  = (b_i, \dots, b_{i+k-1}).
\end{eqnarray}
Again, since $b_i$'s are $\mathbb{F}_p$ numbers, the vector $\vb*{v}_i$ then have  $p^k$ possible values.
The recurrence relation defines a linear mapping from $\vb*{v}_{(i)}$ to  $\vb*{v}_{(i+1)}$, via
\begin{eqnarray}
    \vb*{v}_{i+1} = \vb*{M} . \vb*{v}_{(i)},
\end{eqnarray}
where 
\begin{equation}
   \vb*{M} = 
   \begin{pmatrix}
       -n_k^{-1}n_{k-1}& -n_k^{-1}n_{k-2}&-n_k^{-1}n_{k-3}&\cdots&-n_k^{-1}n_{0}\\
       1& 0 &0 &\cdots & 0\\
       0 & 1 &0&0 & \cdots\\
       0&0&\ddots & 0& 0\\
       0 &  \cdots  & 0&1 & 0
   \end{pmatrix}.
\end{equation}
We note that $n_0$ and $n_k$ are non-zero by definition, otherwise this model reduces to a model with $k$-site $Z$ interactions instead of $k+1$.
Then we have 
\begin{equation}
    \det \vb*{M} = (-1)^{k+1} n_k^{-1}n_{0} \ne 0.
\end{equation}
That is, $\vb*{M}$ defines a one-to-one, onto (i.e. bijection) of the preimage to itself. And this preimage is a finite set $p^k$ elements. 

As such, the map $\vb*{M}$ is a \textit{permutation}. 
Starting from any vector $\vb*{v}$, the sequence 
\begin{equation}
    \vb*{v}, \vb*{M}.\vb*{v}, \vb*{M}^2.\vb*{v},\cdots
\end{equation}
is always periodic, and there is some minimal positive integer number $m$ such that 
\begin{equation}
    \vb*{v} =  \vb*{M}^m.\vb*{v} .
\end{equation}
This sequence $\vb*{v}, \vb*{M}.\vb*{v}, \vb*{M}^2.\vb*{v},\cdots , \vb*{M}^{m-1}.\vb*{v}$ is called an \textit{orbital} of the sequence. Having an \textit{orbital}  of length $m$ means the model has a symmetry with a periodical pattern, whose periodicity is $m$,
\begin{equation}\label{eq: sym operator}
\Tilde{X} = \prod_i X_i^{v_1}X_{i+1}^{(Mv)_1}X_{i+2}^{(M^2v)_1} \cdots X_{i+m-1}^{(M^{m-1}v)_1} X_{i+m}^{v_1}\cdots .
\end{equation}
In fact, there are $m$ such symmetries by shifting the $i$.

Note that the set of $p^k$ possible values of $\vb*{v}$ is divided into individual orbits that do not intersect each other, and every element must belong to one and only one orbital. That means, if we have $l$ distinct orbitals with length $m_1,\dots, m_l$, we must have
\begin{equation}
    m_1 \times \cdots \times m_l = p^k.
\end{equation}

In case of open boundary condition, $\vb*{v}_{(1)}$, or the first $k$ elements of $\vb*{b}$, uniquely determine  the kernel $\vb*{b}$, 
and  such a kernel always uniquely exists for any choice of the first $k$ elements. So we have the full $p^k$ symmetries. 
Let us denote the symmetry operator corresponding to $\vb*{v}$ as $\tilde{X}_{\vb*{v}}$.
Note that $\vb*{v}$ forms a linear vector space: $\vb*{v}$ +  $\vb*{v}'$ and $c\vb*{v}$ still belong to the vector space, and so do the symmetry operators $\tilde{X}_{\vb*{v}}$'s.
Therefore, they symmetries form a group structure of $(\mathbb{Z}_p)^k$

In the thermal dynamics limit (infinitely long chain), the same logic applies if we determines the $b_i$'s to the left of some ``initial condition'' chosen in the middle of the chain. We have the full $(\mathbb{Z}_p)^k$ symmetries. 

In case of periodical boundary condition, when the lattice site number $L$ is multiples of certain length of the orbital(s) of $\vb*{M}$, then the system has symmetries when $\vb*{v}_{(1)}$ belongs to the orbital(s).

Hence we come to the conclusion
\begin{theorem}
Consider the spin chain of qudit $Z$ of $\mathbb{Z}_p$, where $p$ is a prime number. 
In the thermaldynamics limit, the translational symmetric, $k+1$-site interacting Hamiltonian,
\begin{equation}
    \mathcal{H} = \sum_i Z_i^{n_0}Z_{i+1}^{n_1}\cdots Z_{i+k}^{n_k} + X
\end{equation}
has the following properties:
\begin{enumerate}
    \item Given any term $X_i^{b_i}X_{i+1}^{b_{i+1}} \cdots X_{i+k-1}^{b_{i+k-1}}$ specified by $k$ numbers $b_i, \cdots, b_{i+k-1}$ in finite field $\mathbb{F}_p$, there is always one and only one symmetry operators containing this term. 
    \item There are in total $p^k$ symmetries operators, forming the  group structure $(\mathbb{Z}_p)^k$.
    \item Every one of the symmetry operator is periodic in space. Let its periodicity be $m_i$, it is then equivalent to $m_i$  symmetry operators (including itself) by lattice translation symmetry. 
    $m_i$ is the length of the corresponding orbital of the map $\vb*{M}$. 
    \item All  $m_i$'s for different periodicity of the translational equivalent class of symmetry operators satisfy $\prod_i m_i = p^k$.
\end{enumerate}

\end{theorem}

\prlsection{Simplest example}
 The simplest example is the BPM with $n_0=n_2=1$ and $n_1=-2$. The corresponding duality has been discussed in the ref \cite{Cao:2024qjj}, which can be understood as a generalized KW duality associated with dipole symmetry $\Z^{\eta}_p\times \Z^{D}_p$:
 \[
 \eta=\prod_{i}X_i, D=\prod_{i}X^i_i.
 \]
By the approach above, we can directly show how the dipole symmetry can be obtained from this BPM. We first consider 
\begin{equation}
   \vb*{M} = 
   \begin{pmatrix}
       2& -1\\
       1& 0 
   \end{pmatrix}.
\end{equation}
It is straightforward to show that 
\begin{equation}
   \vb*{M}^i = 
   \begin{pmatrix}
       i+1& -i\\
       i& 1-i 
   \end{pmatrix}.
\end{equation}
Thus $M^p$ is the identity matrix and the length of each orbital is $p$. Now let us pick two vectors $v^{\dagger}=(1,0)$ and $\tilde{v}^{\dagger}=(0,1)$. Then we can obtain that $(M^i v)_1=i+1$ and $(M^i \tilde{v})_1=i$. The resulting two symmetry operators in the form \eqref{eq: sym operator} are $D$ and $D\eta^{-1}$.

\section{Anomaly of three-site RBPM symmetry}\label{sec:anomaly}
\prlsection{Three-site RBPM}
In this section, we study the detail of the following  three-site RBPM within the field $\mathbb{F}_p$ ($p$ is prime):
\[
\begin{split}
   &\mathcal N_{\text{R}}\ket{\{s_{i}\}}=\frac{1}{p^{\frac{L}{2}}} \sum_{\{\hat{s}_{i}\}}\omega^{\sum_{j,k=1}^{L}s_{j} A_{jk}\hat{s}_{-k}}\ket{\{\hat{s}_{i}\}} 
\end{split}
\]
where 
\[
A_{j j+1}=rs, A_{j j}=-r-s, A_{j j-1}=1
\]
with $r,s\ne 0$. Here, for simplicity, we consider the above RBPM duality that differs from that in the manuscript by a one-site lattice translation, %for the $s$ system and reflection for both the $s$ and $\hat{s}$ systems. Since translation and reflection do not change the energy spectrum, these two RBPMs share
which yields the same anomaly conditions.
This RBPM can be rewritten in the form

\[
\begin{split}
   &\mathcal N_{\text{R}}\ket{\{s_{i}\}}=\frac{1}{p^{\frac{L}{2}}} \sum_{\{\hat{s}_{i}\}}\omega^{\sum_{j,k=1}^{L}s_{-j} [rs\hat{s}_{j-1}-(r+s)\hat{s}_{j}+\hat{s}_{j+1}]}\ket{\{\hat{s}_{i}\}}, 
\end{split}
\]
which we will mainly use in the following discussion.
The corresponding characteristic equation is  $x^2-(r+s) x+rs=(x-r)(x-s)=0$. We will discuss kernel vector and symmetry-twist sector mapping of BPM duality symmetry in two cases respectively. These results are essential for the anomaly proof presented in the next section. 

In the first case where $r=s$, two independent  kernel vectors are $b^1_j=r^{j}$ and $b^2_j=j r^{j}$. Hence the associated $\Z^{\eta_1}_p\times \Z^{\eta_2}_p$ symmetry is generated by $\eta_1=\prod^L_{j=1}X^{r^{j}}_j$  and $\eta_2=\prod^L_{j=1}X^{jr^{j}}_j$. Similarly the dual symmetry is generated by $\hat{\eta}_1=\prod^L_{j=1}\hat{X}^{r^{j}}_j$  and $\hat{\eta}_2=\prod^L_{j=1}\hat{X}^{jr^{j}}_j$. On the closed chain, it is conventional to assume $r^L=1$ and $L=0 ~ (\text{mod}~ p)$ such that $\eta_1=\prod^{-1}_{j=-L}X^{r^{j}}_j$  and $\eta_2=\prod^{-1}_{j=-L}X^{jr^{j}}_j$.

Now let us consider symmetry twisted boundary conditions: $s_{j+L}=s_j+t_1 r^j+t_2 jr^j$ or equivalently $Z_{j+L}=Z_{j}\omega^{t_1 r^j+t_2 jr^j}$. One can organize the Hilbert space into $p^4$ symmetry-twist sectors, labeled by $(u_{1},t_{1},u_{2},t_{2})$. After applying the three-site RBPM transformation, the  system can similarly be classified into symmetry-twist sectors labeled by $(\widehat{u}_{1},\widehat{t}_{1},\widehat{u}_{2},\widehat{t}_{2})$. Our goal is to determine how these sectors are related before and after the three-site RBPM transformation.

The RBMP duality is modified with general boundary conditions:
\[
\begin{split}
   \mathcal N_{\text{R}}\ket{\{s_{i}\}}&=\frac{1}{p^{\frac{L}{2}}} \sum_{\{\hat{s}_{i}\}}\omega^{\sum_{j=1}^{L}s_{-j}( r^2\hat{s}_{j-1}-2r\hat{s}_{j}+\hat{s}_{j+1})+r(t_2-t_1)\hat{s}_{L}+t_1\hat{s}_{1}}\ket{\{\hat{s}_{i}\}} \\
   &=\frac{1}{p^{\frac{L}{2}}} \sum_{\{\hat{s}_{i}\}}\omega^{\sum_{j=1}^{L}\hat{s}_{j}( r^2s_{-j-1}-2rs_{-j}+s_{-j+1})+r(\widehat{t}_2+\widehat{t}_1)s_{-L}-r^2\widehat{t}_1s_{-1}}\ket{\{\hat{s}_{i}\}}.
\end{split}
\]
Let us first consider $\mathcal N_{\text{R}}\times \eta_{1}$ acting on an arbitrary state 
$\ket{\psi} = \sum_{\{s_i\}} \psi_{\{s_i\}} \ket{\{s_i\}}$.
\begin{equation}
\begin{split}
   \mathcal N_{\text{R}}\times \eta_{1}\ket{\psi} &=  \mathcal N_{\text{R}}  \sum_{\{s_i\}} \psi_{\{s_i\}} \ket{\{s_i+r^i\}}\\
    &= \sum_{\{\hat{s}_{i}\},\{s_i\}} \psi_{\{s_i\}}   \omega^{\sum_{j=1}^{L}\hat{s}_{j}( r^2s_{-j-1}-2rs_{-j}+s_{-j+1})+r(\widehat{t}_2+\widehat{t}_1)(s_{-L}+1)-r^2\widehat{t}_1(s_{-1}+r^{-1})}\ket{\{\hat{s}_{i}\}}\\
     &=\omega^{r\widehat{t}_2}\mathcal N_{\text{R}} \ket{\psi}.
\end{split}
\end{equation}
The result implies that for any eigenstate $\ket{ \Psi}$ with
\begin{eqnarray}
    \eta_1 \ket{ \Psi} = \omega^{u_1} \ket{ \Psi} ,
\end{eqnarray}
we have
\begin{eqnarray}
   \omega^{r\widehat{t}_2} \mathcal N_{\text{R}} \ket{ \Psi} = \mathcal N_{\text{R}}\times  \eta_1\ket{ \Psi}=\omega^{u_{1}} \mathcal N_{\text{R}} \ket{ \Psi},
\end{eqnarray}
namely
\begin{eqnarray}
    r\widehat{t}_2 = u_1.
\label{eq:hatt_u}
\end{eqnarray}
On the other hand, one can also consider  $\widehat{\eta}_1\times\mathcal N_{\text{R}}$ acting on a arbitrary state $\ket{\psi}$,
\begin{equation}
\begin{split}
    \widehat{\eta}_1\times \mathcal N_{\text{R}}  \ket{\psi}
    &= \widehat{\eta}_1\sum_{\{\hat{s}_{i}\},\{s_i\}} \psi_{\{s_i\}}  \omega^{\sum_{j=1}^{L}s_{-j}( r^2\hat{s}_{j-1}-2r\hat{s}_{j}+\hat{s}_{j+1})+r(t_2-t_1)\hat{s}_{L}+t_1\hat{s}_{1}}\ket{\{\hat{s}_{i}\}}\\ 
     &= \sum_{\{\hat{s}_{i}\},\{s_i\}} \psi_{\{s_i\}}  \omega^{\sum_{j=1}^{L}s_{-j}( r^2\hat{s}_{j-1}-2r\hat{s}_{j}+\hat{s}_{j+1})+r(t_2-t_1)\hat{s}_{L}+t_1\hat{s}_{1}}\ket{\{\hat{s}_{i}+r^i\}}\\
      &=\sum_{\{\hat{s}_{i}\},\{s_i\}} \psi_{\{s_i\}}  \omega^{\sum_{j=1}^{L}s_{-j}( r^2\hat{s}_{j-1}-2r\hat{s}_{j}+\hat{s}_{j+1})+r(t_2-t_1)(\hat{s}_{L}-r^L)+t_1(\hat{s}_{1}-r)}\ket{\{\hat{s}_{i}\}}\\
     &=\omega^{-rt_2}\mathcal N_{\text{R}} \ket{\psi}.
\end{split}
\end{equation}
This shows that the dual $\Z^{\eta_1}_p$-symmetry charge sector after RBPM transformation is identified with the twisted sector before this transformation, 
\begin{eqnarray}
\label{eq:hatut}
\widehat{u}_1 = -rt_2.
\end{eqnarray}
By similar calculations, one can determine the correspondence between the remaining sectors:
\[
 u_2=r\widehat{t}_1, \quad \widehat{u}_2 = -rt_1.
\]
In summary, we have the following map of the symmetry-twist sectors:
\[\label{eq: symtwist-1}
(\widehat{u}_{1},\widehat{t}_{1},\widehat{u}_{2},\widehat{t}_{2})=(-rt_2,r^{-1}u_2,-rt_1,r^{-1}u_1).
\]

In the second case where $r\ne s$,  two independent  kernel vectors are $b^1_j=r^{j}$ and $b^2_j= s^{j}$. This case only happens when $p>2$. Hence the associated invertible symmetry is generated by $\eta_1=\prod^n_{j=1}X^{r^{j}}$  and $\eta_2=\prod^n_{j=1}X^{s^{j}}$. Similarly the dual symmetry is generated by $\hat{\eta}_1=\prod^L_{j=1}\hat{X}^{r^{j}}_j$  and $\hat{\eta}_2=\prod^L_{j=1}\hat{X}^{s^{j}}_j$. On the closed chain, it is conventional to assume $r^L=1$ and $s^L=0 ~ (\text{mod}~ p)$ such that $\eta_1=\prod^{-1}_{j=-L}X^{r^{j}}_j$  and $\eta_2=\prod^{-1}_{j=-L}X^{sr^{j}}_j$.

By the same procedure, we also consider symmetry twisted boundary conditions: $s_{j+L}=s_j+t_1 r^j+t_2 s^j$ or equivalently $Z_{j+L}=Z_{j}\omega^{t_1 r^j+t_2 s^j}$. One can organize the Hilbert space into $p^4$ symmetry-twist sectors, labeled by $(u_{1},t_{1},u_{2},t_{2})$. Similarly, the symmetry-twist sectors of spins after the three-site RBPM transformation is labeled by $(\widehat{u}_{1},\widehat{t}_{1},\widehat{u}_{2},\widehat{t}_{2})$.
The RBMP duality is modified with general boundary conditions:
\[
\begin{split}
   \mathcal N_{\text{R}}\ket{\{s_{i}\}}&=\frac{1}{p^{\frac{L}{2}}} \sum_{\{\hat{s}_{i}\}}\omega^{\sum_{j=1}^{L}s_{-j}( r^2\hat{s}_{j-1}-2r\hat{s}_{j}+\hat{s}_{j+1})+rs(-t_1r^{-1}-t_2s^{-1})\hat{s}_{L}+(t_1+t_2)\hat{s}_{1}}\ket{\{\hat{s}_{i}\}} \\
   &=\frac{1}{p^{\frac{L}{2}}} \sum_{\{\hat{s}_{i}\}}\omega^{\sum_{j=1}^{L}\hat{s}_{j}( r^2s_{-j-1}-2rs_{-j}+s_{-j+1})+(s\widehat{t}_2+r\widehat{t}_1)s_{-L}-rs(\widehat{t}_1+\widehat{t}_2)s_{-1}}\ket{\{\hat{s}_{i}\}}.
\end{split}
\]
By the same method, we can identify the mapping of symmetry-twist sector:
\[\label{eq: symtwist-2}
(\widehat{u}_{1},\widehat{t}_{1},\widehat{u}_{2},\widehat{t}_{2})=((s-r)t_1,(r-s)^{-1}u_1,(r-s)t_2,(s-r)^{-1}u_2).
\]

\prlsection{Proof of the anomaly} Now let us discuss the anomaly conditions of the  three-site RBPM symmetry by checking whether there is a unique gapped phase invariant under such duality symmetry. A theory with an anomaly-free symmetry in the low energy should be uniquely gapped. Therefore, we will find an anomaly if we cannot find a gapped phase with a unique ground state that is invariant under the RBPM transformation with a certain $p$. 
The anomaly imposes a nonperturbative constraint on the self-dual theories that they must have continuous or first-order phase transitions.

Let us begin with the case where $r=s$. Now suppose we have a theory with both invertible $\Z^{\eta_1}_p\times \Z^{\eta_2}_p$ symmetry
and the non-invertible three-site RBPM symmetry. We will prove the anomaly of the three-site RBPM symmetry by contradiction. Because   $\Z^{\eta_1}_p\times \Z^{\eta_2}_p$ is onsite, the symmetric theory is compatible with a gapped phase with one ground state, i.e. SPT phase. %We exclude the trivial phase because it is mapped to a SSB phase under the three-site RBPM  transformation and therefore does not have a three-site RBPM symmetry. 
The SPT phase of onsite $\Z^{\eta_1}_p\times \Z^{\eta_2}_p$ symmetry is classified by the group cohomology $H^{2}(\Z^{\eta_1}_p\times \Z^{\eta_2}_p, U(1))=\Z_p$. A simple example is the stabilizer Hamiltonian:
\[\label{eq:SPT Hal}
\mathcal H_{\text{SPT-}k}=-\sum_{i}\sum^{p-1}_{m=0} [(Z^{r^{-2i+2}}_{i-1}Z^{-r^{-2i+1}}_i)^kX_i (Z^{-r^{-2i+1}}_iZ^{r^{-2i}}_{i+1})^k]^m,
\]
where level $k\in \Z_p$ corresponds to different SPT classes. The SPT Hamiltonian is constructed from the trivial-phase Hamiltonian which has $k=0$
\[
\mathcal H_{\text{triv}}=-\sum_i\sum^{p-1}_{m=0}(X_i)^m,
\]
by decorated domain wall construction
\[
\mathcal H_{\text{SPT-}k}=T_D^{k}\mathcal H_{\text{triv}} T_D^{-k},\quad T_{D}=\prod_{i=1}^{L} CZ^{r^{-2i+2}}_{i-1,i}CZ^{-r^{-2i+1}}_{i,i}.
\]
Here the transformation of  control-$Z$ gate on Pauli operators is given by
\[
 CZ_{i,j}=\frac{1}{p}\sum_{\alpha,\beta=1}^{N}\omega^{-\alpha\beta}Z_{i}^{\alpha}Z_{j}^{\beta}: \quad X_{i}\to X_{i}Z_{j},\ X_{j}\to X_{j}Z_{i}.
\]
In particular, when $r^2=1~ \text{mod}~ p$ this Hamiltonian also preserves the translation symmetry.

Since all the terms in the SPT  Hamiltonian~\eqref{eq:SPT Hal} commute with each other, the ground state satisfies
\begin{equation}
    (Z^{r^{-2i+2}}_{i-1}Z^{-r^{-2i+1}}_i)^kX_i (Z^{-r^{-2i+1}}_iZ^{r^{-2i}}_{i+1})^k\ket{\text{G.S.}}=\ket{\text{G.S.}}.
\end{equation} 
The SPT phases with different $k$ are characterized by the charges of ground state $\ket{\text{G.S.}}_{t_1,t_2}$ with twisted boundary conditions labelled by $(t_1,t_2)$~\cite{Li:2022jbf,PhysRevB.106.224420}. 
It is straightforward to calculate the charges of symmetry operator $\eta_1,\eta_2$ on ground state $\ket{\text{G.S.}}_{t_1,t_2}$
\[\label{eq:SPT-GS}
\begin{split}
&(\prod^L_{i=1}X^{r^i}_{i})\ket{\text{G.S.}}_{t_1,t_2}=\prod^L_{i=1} (Z^{r^{-2i+2}}_{i-1}Z^{-2r^{-2i+1}}_iZ^{r^{-2i}}_{i+1})^{-kr^i}\ket{\text{G.S.}}_{t_1,t_2}=(Z^{-r}_0Z_1 Z^{r}_L Z^{-1}_{L+1})^k\ket{\text{G.S.}}_{t_1,t_2}=\omega^{-krt_2}\ket{\text{G.S.}}_{t_1,t_2},\\
&(\prod^L_{i=1}X_i^{ir^i})\ket{\text{G.S.}}_{t_1,t_2}=\prod^L_{i=1} (Z^{r^{-2i+2}}_{i-1}Z^{-2r^{-2i+1}}_iZ^{r^{-2i}}_{i+1})^{-kir^i}\ket{\text{G.S.}}_{t_1,t_2}=(Z^{-r}_0 Z^r_L)^k\ket{\text{G.S.}}_{t_1,t_2} =\omega^{rkt_1}\ket{\text{G.S.}}_{t_1,t_2},
\end{split}
\]
where we used the twist boundary condition of Pauli operators in the last equality in each equation.
Therefore the ground state $\ket{\text{G.S.}}_{t_1,t_2}$ is in the symmetry-twist sector labelled by 
\begin{equation}\label{eq:twistbefore}
    (u_1=-rkt_2, t_1, u_2=rkt_1, t_2).
\end{equation}
If the ground states after the three-site BPM transformation do not stay in the same sector, the SPT phase is not invariant under this transformation. If for given $p$ every $k$ we cannot find an SPT that is invariant under the RBPM transformation, then the symmetry is anomalous.

Now let us check whether there is an SPT phase invariant under three-site BPM transformation. The dual Hamiltonian is 
\[\label{eq:dual-SPT-Hal}
\begin{split}
\mathcal H'_{k}=-\sum_{i}\sum^p_{m=1} [\hat{Z}^{r^2}_{-i-1}\hat{Z}^{-r}_{-i}\hat{X}^{-kr^{-2i}}_{-i}\hat{Z}^{-r}_{-i}\hat{Z}_{-i+1}]^m=-\sum_{i}\sum^p_{m=1} [\hat{Z}^{r^2}_{i-1}\hat{Z}^{-r}_{i}\hat{X}^{-kr^{2i}}_{i}\hat{Z}^{-r}_{i}Z_{i+1}]^m.
\end{split}
\]
In the dual systems, we label twist boundary conditions of dual systems using $\widehat{t}_{1},\widehat{t}_{2}$. Due to the mapping of symmetry-twist sectors under RBPM in Eq.\eqref{eq: symtwist-1} and the symmetry-twist sectors of ground state before duality, we can show that the dual ground state $\ket{\text{G.S.}'}_{\widehat{t}_{1},\widehat{t}_{2}}$ is in the symmetry-twist sector labelled by
\begin{equation}\label{eq:twistafter}
    (\widehat{u}_1=rk^{-1}\widehat{t}_2, \widehat{t}_1, \widehat{u}_2=-r k^{-1}\widehat{t}_1, \widehat{t}_2).
\end{equation}
Therefore, if there is an SPT phase invariant under RBPM transformation, the ground state charges of \eqref{eq:dual-SPT-Hal} and  \eqref{eq:SPT Hal} under the same twisted boundary condition should be consistent with each other, that is $k^2=-1 ~(\text{mod}~ p)$
Thus  $k$ should satisfy $k^2=-1 ~ \text{mod}~ p$, i.e., $-1$ is a quadratic residue modulo $p$, which is the necessary anomaly-free condition for RBPM symmetry.

Indeed, this condition is also a sufficient anomaly-free condition. This is because the dual Hamiltonian \eqref{eq:dual-SPT-Hal} is the same as the Hamiltonian  \eqref{eq:SPT Hal} in this case.  For each $m\in \Z_p$ and site $i$, there exist a unique $j_{m,i}\in \Z_p$ satisfying $-k r^{2i}j_{m,i}=m ~ \text{mod}~ p$. As a result, we have the 
\[
\begin{split}
\mathcal H'_{k}&=-\sum_{i}\sum^p_{m=1} [\hat{Z}^{r^2}_{i-1}\hat{Z}^{-r}_{i}\hat{X}^{-kr^{2i}}_{i}\hat{Z}^{-r}_{i}\hat{Z}_{i+1}]^{j_{m,i}}\\&=-\sum^L_{i=1}\sum^N_{m=1} \hat{Z}^{r^2j_{m,i}}_{i-1}\hat{Z}^{-rj_{m,i}}_i \hat{X}^m_i \hat{Z}^{-rj_{m,i}}\hat{Z}^{j_{m,i}}_{i+1}\\
&=-\sum^L_{i=1}\sum^N_{m=1} \hat{Z}^{kmr^{-2i+2}}_{i-1}\hat{Z}^{-kmr^{-2i+1}}_i \hat{X}^m_i \hat{Z}^{-kmr^{-2i+1}}\hat{Z}^{kmr^{-2i}}_{i+1}=H_{\text{SPT-}k},
\end{split}
\]
where in the last equation, we use the fact that $j_{m_i}=-k^2 j_{m,i}=kmr^{-2i}  ~ \text{mod}~ p$.

Furthermore, we can also apply the same method to discuss the anomaly of RBPM symmetry when $r\ne s$. The SPT phase projected by $\Z^{\eta_1}_p\times \Z^{\eta_2}_p$ is the stabilizer Hamiltonian:
\[\label{eq:SPT Hal-1}
\mathcal H_{\text{SPT-}k}=-\sum_{i}\sum^p_{m=1} [(Z^{(rs)^{-i+1}}_{i-1}Z^{-\frac{(r+s)}{2}(rs)^{-i}}_i)^kX_i (Z^{-\frac{(r+s)}{2}(rs)^{-i}}_iZ^{(rs)^{-i}}_{i+1})^k]^m,
\]
where level $k\in \Z_p$ corresponds to different classes. The SPT Hamiltonian is constructed from the trivial-phase Hamiltonian
\[
\mathcal H_{\text{triv}}=-\sum_{i}\sum^p_{m=1}(X_i)^m,
\]
by decorated domain wall construction
\[
\mathcal H_{\text{SPT-}k}=T_D^{k}\mathcal H_{\text{triv}} T_D^{-k},\quad T_{D}=\prod_{i=1}^{L} CZ^{(rs)^{-i+1}}_{i-1,i}CZ^{-\frac{(r+s)}{2}(rs)^{-i}}_{i,i}.
\]
In particular, when $rs=1~ \text{mod}~ p$ this Hamiltonian also preserves the translation symmetry.

It is straightforward to calculate the charges of symmetry operator $\eta_1,\eta_2$ on ground state $\ket{\text{G.S.}}_{t_1,t_2}$
\[\label{eq:SPT-GS-1}
\begin{split}
&(\prod^L_{i=1}X^{r^i}_{i})\ket{\text{G.S.}}_{t_1,t_2}=\prod^L_{i=1} (Z^{(rs)^{-i+1}}_{i-1}Z^{-(r+s)(rs)^{-i}}_i Z^{(rs)^{-i}}_{i+1})^{-kr^i}\ket{\text{G.S.}}_{t_1,t_2}=(Z^{-r}_0Z_1 Z^{r}_L Z^{-1}_{L+1})^k\ket{\text{G.S.}}_{t_1,t_2}=\omega^{k(r-s)t_2}\ket{\text{G.S.}}_{t_1,t_2},\\
&(\prod^L_{i=1}X^{s^i}_{i})\ket{\text{G.S.}}_{t_1,t_2}=\prod^L_{i=1} (Z^{r^{-2i+2}}_{i-1}Z^{-2r^{-2i+1}}_iZ^{r^{-2i}}_{i+1})^{-kir^i}\ket{\text{G.S.}}_{t_1,t_2}=(Z^{-s}_0Z_1 Z^{s}_L Z^{-1}_{L+1})^k\ket{\text{G.S.}}_{t_1,t_2} =\omega^{k(s-r)t_1}\ket{\text{G.S.}}_{t_1,t_2}.
\end{split}
\]
This implies the ground state $\ket{\text{G.S.}}_{t_1,t_2}$ is in the symmetry-twist sector labelled by 
\begin{equation}\label{eq:twistbefore-1}
    (u_1=k(r-s)t_2, t_1, u_2=k(s-r)t_1, t_2).
\end{equation}
The dual Hamiltonian is 
\[\label{eq:dual-SPT-Hal-1}
\begin{split}
\mathcal H'_{k}=-\sum_{i}\sum^p_{m=1} [\hat{Z}^{rs}_{-i-1}\hat{Z}^{-\frac{(r+s)}{2}}_{-i}\hat{X}^{-k(sr)^{-i}}_{-i}\hat{Z}^{-\frac{(r+s)}{2}}_{-i}\hat{Z}_{-i+1}]^m=-\sum_{i}\sum^p_{m=1} [\hat{Z}^{rs}_{i-1}\hat{Z}^{-\frac{(r+s)}{2}}_{i}\hat{X}^{-k(sr)^{i}}_{i}\hat{Z}^{-\frac{(r+s)}{2}}_{i}\hat{Z}_{-i+1}]^m.
\end{split}
\]
  Due to the mapping of symmetry-twist sectors under RBPM in Eq.\eqref{eq: symtwist-2} and the symmetry-twist sectors of ground state before duality, we can show that the ground state $\ket{\text{G.S.}'}_{\widehat{t}_{1},\widehat{t}_{2}}$ is in the symmetry-twist sector labeled by
\begin{equation}\label{eq:twistafter-1}
    (\widehat{u}_1=k^{-1}(s-r)\widehat{t}_2, \widehat{t}_1, \widehat{u}_2= k^{-1}(r-s)\widehat{t}_1, \widehat{t}_2).
\end{equation}
Thus if there is an SPT phase invariant under RBPM transformation, we have $k^2=-1 ~(\text{mod}~ p)$, i.e., $-1$ is a quadratic residue modulo $p$, which is the necessary anomaly-free condition for RBPM symmetry.

Indeed, this condition is also a sufficient anomaly-free condition for both two cases. This is because the dual Hamiltonian is the same as the Hamiltonian when this condition is satisfied. We also note that if we further choose $r=s$, the dual Hamiltonian \eqref{eq:dual-SPT-Hal-1} and the Hamiltonian  \eqref{eq:SPT Hal-1} will become the dual Hamiltonian \eqref{eq:dual-SPT-Hal} and the Hamiltonian  \eqref{eq:SPT Hal}. Thus we only need to prove that the dual Hamiltonian \eqref{eq:dual-SPT-Hal-1} is the same as the Hamiltonian  \eqref{eq:SPT Hal-1} when $k^2=-1 ~ \text{mod}~ p$.   For each $m\in \Z_p$ and site $i$, there exist a unique $j_{m,i}\in \Z_p$ satisfying $-k (rs)^{i}j_{m,i}=m ~ \text{mod}~ p$. As a result, we have the 
\[
\begin{split}
\mathcal H'_{k}&=-\sum_{i}\sum^p_{m=1} [\hat{Z}^{rs}_{i-1}\hat{Z}^{-\frac{(r+s)}{2}}_{i}\hat{X}^{-k(sr)^{i}}_{i}\hat{Z}^{-\frac{(r+s)}{2}}_{i}\hat{Z}_{-i+1}]^{j_m}\\&=-\sum^L_{i=1}\sum^N_{m=1} \hat{Z}^{rsj_{m,i}}_{i-1}\hat{Z}^{-\frac{(r+s)}{2}j_{m,i}}_i \hat{X}^m_i \hat{Z}^{-\frac{(r+s)}{2}j_{m,i}}\hat{Z}^{j_{m,i}}_{i+1}\\
&=-\sum^L_{i=1}\sum^N_{m=1} \hat{Z}^{km(rs)^{-i+1}}_{i-1}\hat{Z}^{-km\frac{(r+s)}{2}(rs)^{-i}}_i \hat{X}^m_i \hat{Z}^{-km\frac{(r+s)}{2}(rs)^{-i}}\hat{Z}^{km(rs)^{-i}}_{i+1}=H_{\text{SPT-}k},
\end{split}
\]
where in the last equation, we use the fact that $j_{m_i}=-k^2 j_{m,i}=km(rs)^{-i}  ~ \text{mod}~ p$. 

In summary, when $-1$ is a quadratic residue modulo $p$, the three-site RBPM symmetry is anomaly-free and there exists a unique gapped phase invariant by this non-invertible symmetry; otherwise it is an anomalous symmetry.\\

\prlsection{Examples with $p=3$}
The smallest integer with anomalous RBPM is $p=3$. In general, there are three cases: 1. $r=s=1$, 2. $r=-s=-1$, 3. $r=s=-1$. 

In the first case where $r=s=1$, the invertible symmetry is generated by $\eta_1=\prod_j X_j$ and $\eta_2=\prod_j X^j_j$ which is dipole symmetry. This anomalous RBPM duality symmetry can help us understand the phase transition in the model:
\begin{equation}\label{eq:TF HAL}
\begin{split}
    \mathcal H_{\text{dipole-Ising}}
    &=\sum_{i=1}^{L}(\cos\theta Z_{i-1}(Z_{i}^{\dagger})^2Z_{i+1}+\sin\theta X_{i})+\text{(h.c.)}.
    \end{split}
\end{equation}
The phase diagram of this model has been discussed in the reference \cite{Cao:2024qjj}. In particular, when $\theta=\pi/4$, the system is gapless with $c=2$ while  when $\theta=5\pi/4$ the system is in a first-order phase transition with dengerate ground states.  Both parameters are self-dual under RBPM whose phase-transition feature is ensured by the anomalous duality symmetries.

%In the case where $r=s=1$, the invertible symmetry is generated by $\eta_1=\prod_j X_j$ and $\eta_2=\prod_j X^j_j$ which is dipole symmetry.% This anomaly can help us understand the phase transition in the model:
%\begin{equation}\label{eq:TF HAL}
%\begin{split}
%    H_{\text{dipole-Ising}}
%    &=\sum_{i=1}^{L}(\cos\theta Z_{i-1}(Z_{i}^{\dagger})^2Z_{i+1}+\sin\theta X_{i})+\text{(h.c.)},
%    \end{split}
%\end{equation}
%The phase diagram of this model has been discussed in the reference. It has a second-order phase transition at $\theta=0.25\pi$ and a first order phase transition at $\theta=1.25\pi$, which are self-dual ponits under RBPM duality. Hence these two phase-transition can be understood by the anomalous RBPM duality symmetries.

The second case has $r=-s=1$, where the invertible symmetry is generated by $\eta_1=\prod_j X_j$ and $\eta_2=\prod_j X^{(-1)^j }_j$ %or equivalently $\eta_{o}=\prod_j X_{2j-1}$ and $\eta_{e}=\prod_j X_{2j}$ which are spin flip on odd and even sites. 
The corresponding model is 
\[
\mathcal H_{\text{3-state potts}}=\cos\theta\sum_i Z_{i-1}Z^{\dagger}_{i+1}+\sin\theta\sum_i  X_{i}+\text{(h.c.)}.
\]
which are two decoupled 3-state potts models. The self-dual point under RBPM duality is at $\theta=\pi/4$ and $\theta=5\pi/4$. The low energy theory is two decoupled $U(1)_6$ CFTs when $\theta=\pi/4$  and is two decoupled 3-state potts CFTs when $\theta=5 \pi/4$ \cite{lahtinen2021quantum}. The gapless feature of both self-dual points is guaranteed by the anomalous RBPM duality symmetry.

In the final case with $r=s=-1$ which we show in the main text, the invertible symmetry is generated by $\eta_1=\prod_j X^{(-1)^j}_j$ and $\eta_2=\prod_j X^{(-1)^j j}_j$.%, which we denote as "staggered dipole symmetry".
The generalized qudit Ising model is given by
\begin{equation}
\begin{split}
    \mathcal H_{\text{sdipole-Ising}}
    &=\sum_{i=1}^{L}(\cos\theta Z_{i-1}Z_{i}^2Z_{i+1}+\sin\theta X_{i})+\text{(h.c.)}.
    \end{split}
\end{equation}

This model can be mapped to the quantum torus chain 
\begin{equation}\label{eq:quan torus Hal}
    \mathcal H_{\text{QT}}=\sum_{i=1}^{L}(\cos\theta X_{i-1}^{\dagger}X_{i}+\sin\theta Z_{i-1}^{\dagger} Z_i)+\text{(h.c.)},
\end{equation}
 by the BPM duality with matrix $A_{ii}=(-1)^i, A_{ii+1}=(-1)^{i+1}$. That is
 \[
\begin{split}
         \mathcal N_1\ket{\{s_{i}\}}=\frac{1}{3^{\frac{L}{2}}} \sum_{\{\hat{s}_{i}\}}\omega^{\sum_{k=1}^{L}(-1)^k (s_{k-1}+s_{k})\hat{s}_{k}}\ket{\{\hat{s}_{i}\}}.
\end{split}
\]
 This BPM duality induce the transformation as follows:
  \[
\begin{split}
         \mathcal N_1 X_j\ket{\{s_{i}\}}=\frac{1}{3^{\frac{L}{2}}} \sum_{\{\hat{s}_{i}\}}\omega^{\sum_{k=1}^{L}(-1)^k (s_{k-1}+s_{k})\hat{s}_{k}+(-1)^j\hat{s}_{j}+(-1)^{j+1}\hat{s}_{j+1}}\ket{\{\hat{s}_{i}\}}=(\hat{Z}_j\hat{Z}^{\dagger}_{j+1})^{(-1)^j}\mathcal N_1\ket{\{s_{i}\}},\\
          \hat{X}_j \mathcal N_1 \ket{\{s_{i}\}}=\frac{1}{3^{\frac{L}{2}}} \sum_{\{\hat{s}_{i}\}}\omega^{\sum_{k=1}^{L}(-1)^k (s_{k-1}+s_{k})\hat{s}_{k}-(-1)^j (s_{j-1}+s_{j})}\ket{\{\hat{s}_{i}\}}=\mathcal N_1 (Z_{j-1}Z_j)^{(-1)^{j-1}}\ket{\{s_{i}\}}.
\end{split}
\]
 The  $\Z_3^{\eta_2}$  symmetry becomes a $\Z^Z_3$ symmetry generated by $\prod^L_{i=1}Z_i$ and we have a new quantum $\mathbb Z_3^X$ symmetry generated by $\prod_{i=1}^LX_i$.  The phase diagram of the Hamiltonian \eqref{eq:quan torus Hal}, known as the quantum torus chain, has been determined numerically in \cite{PhysRevB.86.134430,PhysRevB.104.045151}.  The results can be summarized as follows:
\begin{enumerate}
    %\item The phase diagram exhibits a symmetric pattern across the line of $\theta=0.25\pi$, $\theta=1.25\pi$, at which points the model is invariant under the global Hadamard gate $U^{H}: X_{i}\to Z_i^{\dagger},Z_i\to X_{i},\forall i$. 
    \item Around $\theta=\pi/4$, the model exhibits an extended gapless phase with central charge $c=2$, covering $\theta\in (-0.1\pi,0)\cup(0,0.5\pi)\cup(0.5\pi,0.6\pi)$. At two particular points $\theta=0,0.5\pi$ there are first-order phase transitions with exponentially large ground state degeneracy.
    \item At $\theta=5\pi/4$, a first-order phase transition separates two distinct symmetry-broken phases:  the $\Z^Z_3$ SSB phase with nonzero $\langle X_i \rangle$ when $\theta\in(0.6\pi,1.25\pi)$ and the $\Z^X_3$ SSB phase with nonzero $\langle Z_i \rangle$ when $\theta\in (1.25\pi,1.9\pi)$. The first-order phase transitions at $\theta=0.6\pi, 1.9\pi$ separate gapped SSB phases and gapless phases. 
\end{enumerate} 
As the BPM duality preserves both the structure of the phase diagram and the central charge of the gapless regions, we can deduce the phase diagram of the staggered dipole Ising model as follows:
\begin{enumerate}
    \item Around $\theta=\pi/4$, the system is in the gapless phases with $c=2$ in the region $\theta\in (-0.1\pi,0)\cup(0,0.5\pi)\cup(0.5\pi,0.6\pi)$. These phases are separated by first-order phase transitions at $\theta=0,0.5\pi$, where the Hamiltonian is dominated by  $Z_{i-1}Z_{i}^2Z_{i+1}+\text{(h.c.)}$ and $X_i+\text{(h.c.)}$
    respectively. At both transition points, the system exhibits an exponentially large ground-state degeneracy.
    \item At $\theta=5\pi/4$, the system undergoes a first-order transition separating a trivially gapped phase from a staggered- dipole SSB phase.  For $\theta\in(1.25\pi,1.9\pi)$, the Hamiltonian is dominated by $-X_i+\text{(h.c.)}$ and the system realizes the trivially gapped phase.  For $\theta\in (0.6\pi,1.25\pi)$, the Hamiltonian is dominated by  $-Z_{i-1}Z_{i}^2Z_{i+1}+\text{(h.c.)}$. There are nine ground states when $L\to \infty$ (with the sequence of $L\equiv 0\text{ mod }3$) with nonzero $\langle Z_{i-1}Z_i\rangle$ and $\langle Z_i\rangle$, leading to t $\Z^{\eta_1}_3\times \Z^{\eta_2}_3$ SSB phase. There are also first-order phase transitions at $\theta=0.6\pi, 1.9\pi$ between different gapped phases and gapless phases.   
\end{enumerate}

\end{document}